\documentclass[a4paper,fleqn,usenatbib]{mnras}

\usepackage{newtxtext,newtxmath}

\usepackage[T1]{fontenc}
\usepackage{ae,aecompl}

\usepackage{graphicx}	
\usepackage{amsmath}	
\usepackage{amssymb}	
\usepackage{tabu}  
\usepackage{gensymb}  
\usepackage{subcaption}  
\title[Gravitational waves from transient mountains]{Transient gravitational waves from pulsar post-glitch recoveries}

\author[G. Yim \& D. I. Jones]{
Garvin Yim\thanks{E-mail: g.yim@soton.ac.uk} and
D. I. Jones
\\
Mathematical Sciences and STAG Research Centre, University of Southampton, Southampton SO17 1BJ, UK
}

\date{Accepted XXX. Received YYY; in original form ZZZ}

\pubyear{2020}

\begin{document}
\label{firstpage}
\pagerange{\pageref{firstpage}--\pageref{lastpage}}
\maketitle

\begin{abstract}
This work explores whether gravitational waves (GWs) from neutron star (NS) mountains can be detected with current 2nd-generation and future 3rd-generation GW detectors. In particular, we focus on a scenario where transient mountains are formed immediately after a NS glitch. In a glitch, a NS's spin frequency abruptly increases and then often exponentially recovers back to, but never quite reaches, the spin frequency prior to the glitch. If the recovery is ascribed to an additional torque due to a transient mountain, we find that GWs from that mountain are marginally-detectable with Advanced LIGO at design sensitivity and is very likely to be detectable for 3rd-generation detectors such as the Einstein Telescope. Using this model, we are able to find analytical expressions for the GW amplitude and its duration in terms of observables. 
\end{abstract}





\section{Introduction}
\label{Introduction}

Traditionally, searches for GWs have been directed towards detecting compact binary coalescences (CBCs), continuous GWs (CWs), burst GWs and stochastic GWs. Of most relevance here are CWs which are GWs with long durations compared to the time spent observing them and are quasi-monochromatic in nature. CBC GWs and CWs have modelled waveforms meaning the method of matched-filtering can be used to detect them. Both burst and stochastic GWs are generally unmodelled making them particularly difficult to detect. Out of these four GW groups, only CBC GWs have been detected so far \citep{abbottetal2019GWTC1}.

In this work, we will be focusing on \textit{transient} CWs from individual NSs \citep[e.g.][]{prixGiampanisMessenger2011}. Transient CWs, a type of long-duration transient GW, are CWs which have durations between the time-scales of CBCs and conventional CWs. Roughly speaking, this means they have durations on the order of hours to weeks, well within the duration of any Advanced Virgo or Advanced LIGO (aLIGO) observation run which is on the order of several months \citep{acerneseetal2015, LSC2015}. 

So far, two types of complementary long-duration transient GW searches have been developed. One from \citet{thraneetal2011}, who took a burst search (looking for unmodelled GWs on the duration of $\mathcal{O}$(seconds)) and extended it to include GWs in our time-scales of interest. The other type came from the creators of the term ``transient CWs'', \citet{prixGiampanisMessenger2011}, who took a conventional (modelled) CW search and shortened the CW duration instead. There are other examples of the latter in \citet{abbottetal2019longringdown}. 

There have been a couple of attempts to use these searches to detect long-duration transient GWs. Most recently, \citet{keiteletal2019} tried to find transient CWs in the post-glitch recoveries of the Crab and Vela pulsars but unfortunately, no transient CWs were detected. \citet{abbottetal2019longringdown} did a similar search but for unmodelled transient GWs after the binary NS merger GW170817, for durations of up to 8.5 days. Again, no transient GWs were detected. 

There have been many other searches conducted by aLIGO, but these were for transient GWs with durations shorter than the time-scales we are interested in here, which includes burst searches. These include transient GWs after magnetar bursts \citep{abbottetal2019magnetarbursts}, during the ringdown (up to 500~s after) of GW170817 \citep{abbottetal2017ringdown}, after long gamma-ray bursts \citep{abbottetal2019gammaraybursts} and those due to NS oscillations after the Vela pulsar glitched in 2006 \citep{abadieetal2011Vela}. There was also an all-sky search within O2 data for transient GWs which were 2 -- 500~s in duration but besides from recovering GW170817, no other transients were detected \citep{abbottetal2019allsky}. 

On the other end of the GW spectrum, the CW spin-down limit for the Crab and Vela pulsars were beaten in 2008 and 2011 respectively \citep{abbottetal2008crabspindown, abadieetal2011velaspindown} with recently updated GW strain upper limits of $h_0^\text{Crab} \lesssim 2 \times 10^{-26}$ and \mbox{$h_0^\text{Vela} \lesssim 2 \times 10^{-25}$} \citep{abbottetal2020upperlimits}. These upper limits were calculated assuming a coherent CW signal for the duration of the observing time. It is perfectly acceptable for transient CWs to have values of $h_0$ larger than these upper limits due to their shorter duration.

It is evident that there are searches being conducted for transient CWs but thus far, there has been little theoretical modelling of how these transient CWs come about. Pulsar glitches, spontaneous step-like increases in a NS's spin frequency, have often been explored as possible triggers. The idea of Ekman pumping after pulsar glitches, which is the bulk movement of fluids due to a tangential force (in the case of a glitching two-component NS, it is the viscous shear force between the crust and core), has been suggested \citep{vanEysdenMelatos2008,bennettvanEysdenMelatos2010,singh2017} as well as polar Alfv\'en waves from magnetar flares \citep{kashiyamaIoka2011}. Additionally, \citet{prixGiampanisMessenger2011} created a toy model proposing the excess superfluid energy from a pulsar glitch could seed a non-axisymmetric deformation leading to transient CW emission. One obstacle for this model is it depends on knowing the lag between the superfluid core and the crust or the superfluid angular velocity, both of which we cannot currently observe. 

Here, we propose another transient CW model which is simple and makes precise and falsifiable predictions. Like other models, it uses pulsar glitches as the trigger and with our model, we are able to naturally explain the post-glitch recovery found immediately after a pulsar glitch. We concern ourselves with only the post-glitch recovery and not the cause of the glitch.

\begin{figure}
	\centering
	\includegraphics[width=\columnwidth]{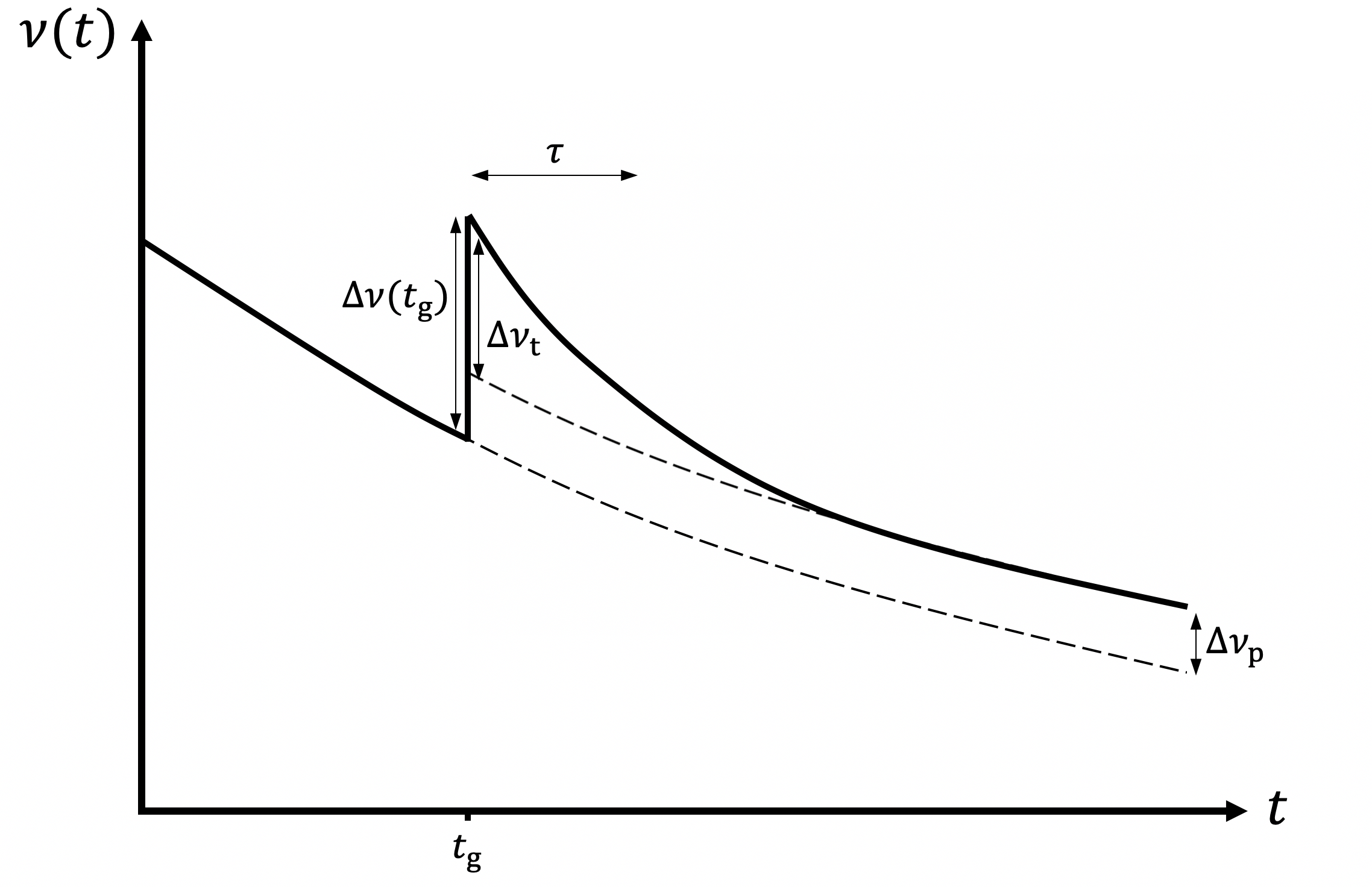} 
	\caption{
		\label{fig:nufunctiont}
		The graph shows how the spin frequency of a NS changes as a function of time when there is a glitch. There are two parts, one being the secular spin-down (shown by the lower dashed line) and the other being the glitch. $\Delta\nu(t_\text{g})$ is the total change in spin frequency at the time of the glitch, which is made up of a permanent part, $\Delta\nu_\text{p}$, and a transient part,~$\Delta\nu_\text{t}$. $\Delta\nu_\text{t}$ exponentially decays away with a time-scale of $\tau$. Also, the permanent change to the time derivative of the spin frequency, $\Delta\dot{\nu}_\text{p}$, has been set to $\Delta\dot{\nu}_\text{p} = 0$ in the above graph, which has been assumed for most of our analysis.
	}
\end{figure}

Fig.~\ref{fig:nufunctiont} shows the usual picture of a pulsar glitch. Tracing from the left, we see the NS's spin frequency, $\nu$, decreases on a secular time-scale until the point of the glitch which occurs at time $t_\text{g}$. The glitch is defined as an instantaneous increase in a NS's spin frequency ($\Delta\nu(t_\text{g}) > 0$) whilst the time derivative of the spin frequency simultaneously decreases ($\Delta\dot{\nu}(t_\text{g}) < 0$). Then, immediately after the glitch is the post-glitch recovery where the spin frequency and its time derivative are observed to be well-modelled by an exponential recovery \citep[e.g.][]{lyneShemarGraham-Smith2000, espinozaetal2011}. Both these parameters generally do not fully recover to pre-glitch values \citep[e.g.][]{haskellMelatos2015}.

In our model, we propose the decrease in $\dot{\nu}$ (i.e. an increase in the spin-down rate, $\Delta|\dot{\nu}|>0$) during the post-glitch recovery is due to an external braking torque caused by a NS ``mountain''. This corresponds to a change to the shape of the NS into a tri-axial ellipsoid (a $l=2$, $m=2$ deformation in the language of spherical harmonics). Along with rotation, there is enough within the model to give the time-varying mass quadrupole moment required for GW emission. The rest of the post-glitch recovery is explained by the mountain slowly dissipating away, on a similar time-scale as the glitch recovery time-scale, $\tau$, leading to a reduction in the braking torque until the spin-down rate recovers. The dissipation could be caused by plastic flow \citep[e.g.][]{baikoChugunov2018} of the NS crust or magnetic diffusion \citep[e.g.][]{ponsVigano2019} but we do not focus on any particular mechanism here. Similarly, we do not focus on the formation mechanism of transient mountains. We only explore the consequences of such mountains. These transient mountains would give rise to transient CWs.

It should be noted that our model is not the usual explanation for the post-glitch recovery which is normally attributed to the model of vortex creep, pioneered by \citet{alparetal1984}. The vortex creep model builds on the vortex unpinning glitch model of \citet{andersonItoh1975} which relies on there being two components to a NS, a normal component and a superfluid component \citep{baymPethickPines1969}. It is generally accepted that large glitches require some superfluid contribution \citep{alparetal1993} but in our simple model, we choose to ignore any internal effects.

In this paper, we take our transient mountain model and calculate whether the post-glitch transient CWs given off would be detectable with current and future GW detectors. Our model is largely motivated by the fact we have observations of pulsar glitches at radio wavelengths, and so we have a set of observables to shape our model. This is particularly topical with the rapid developments seen recently within multi-messenger astronomy. We also aim to interpret and provide physical context to why searches such as \cite{keiteletal2019} should continue in the future.

In Section~\ref{The glitch model from radio astronomy}, we introduce and develop relevant equations from radio pulsar astronomy that describe glitches and the subsequent recovery. In Section~\ref{Maximum ellipticity and gravitational wave strain}, we look at the dynamics of the system which allows us to calculate the ellipticity required to cause the post-glitch recovery. This, in turn, tells us the GW strain achievable. In Section~\ref{Total gravitational wave energy from a glitch}, we find an expression for the total GW energy radiated away by a transient mountain. In Section~\ref{Maximum signal-to-noise ratio}, we relate this GW energy to the signal-to-noise ratio (SNR) achievable in a GW detector. In Section~\ref{Results}, we apply our model to radio data and give a table of results including the SNRs for Crab and Vela glitches for both aLIGO and the Einstein Telescope (ET). In Section~\ref{Discussion}, we discuss the predictions and limitations of our model. Finally, Section~\ref{Conclusion} summarises all the results of this work and concludes with some final remarks on future prospects of this research.

\section{The glitch model from radio astronomy}
\label{The glitch model from radio astronomy}

Timing a pulsar requires accurate modelling of the rotational dynamics of a NS. One well-known property of NSs is that they spin-down on long (secular) time-scales. This is thought to be primarily due to magnetic dipole radiation \citep[e.g.][]{lyneGraham-Smith2012} though searches are being performed to see if CWs could also play a role \citep[e.g.][]{abbottetal2020upperlimits}. This secular spin-down is represented by the lower dashed line in Fig.~\ref{fig:nufunctiont}. Since we are only interested in glitches (which have much shorter time-scales), we can subtract the secular spin-down to leave the change in the spin frequency due to the glitch, $\Delta\nu(t)$, represented by 
\begin{equation}
\label{phenomenologicalModel}
\Delta\nu(t) = \left.
\begin{cases}
0~, & ~~~\text{if } t < t_\text{g} \\
\Delta\nu_\text{p} + \Delta\dot{\nu}_\text{p} \cdot \Delta t + \Delta\nu_\text{t}e^{-\frac{\Delta t}{\tau}}~, & ~~~\text{if } t \ge t_\text{g} 
\end{cases}
\right.
\end{equation}
where $\Delta t = t - t_\text{g}$ is the time elapsed after the glitch with $t_\text{g}$ being the time of the glitch. $\Delta\nu_\text{p}$ refers to the change in spin frequency which is permanent due to the glitch and $\Delta\nu_\text{t}$ is the same but refers to the transient part, i.e. the change in the spin frequency which fully recovers on time-scales much larger than the recovery time-scale of the glitch, $\tau$. The product $\Delta\dot{\nu}_\text{p} \cdot \Delta t$ represents the contribution to $\Delta\nu(t)$ due to a permanent change in the spin-down rate caused by the glitch, $\Delta\dot{\nu}_\text{p}$. From this formulation, it is seen at $t = t_\text{g}$, we get the relation
\begin{equation}
\label{nuSum}
\Delta\nu(t_\text{g}) = \Delta\nu_\text{p} + \Delta\nu_\text{t}~.
\end{equation} 

We can then differentiate Equation (\ref{phenomenologicalModel}) to get the time derivative of the spin frequency, $\Delta\dot{\nu}(t)$, and in the regime of $t \ge t_\text{g}$, we have
\begin{equation}
\Delta\dot{\nu}(t) = \Delta\dot{\nu}_\text{p} - \frac{\Delta\nu_\text{t}}{\tau}   e^{-\frac{\Delta t}{\tau}}~.
\end{equation} 

From this, we associate the coefficient of the time-dependent term to be equal to the transient change in the spin-down rate, $\Delta\dot{\nu}_\text{t}$, which is given by
\begin{equation}
\label{def: delta dot nu t}
\Delta\dot{\nu}_\text{t} \equiv - \frac{\Delta\nu_\text{t}}{\tau}~,
\end{equation}
such that $\Delta\dot{\nu}(t) = \Delta\dot{\nu}_\text{p} + \Delta\dot{\nu}_\text{t} e^{-\frac{\Delta t}{\tau}} = \Delta\dot{\nu}_\text{p} + \Delta\dot{\nu}_\text{t}(t)$. There is a slight subtlety in the notation here since we have explicitly used parentheses to represent a time-dependence, e.g. $\Delta\dot{\nu}_\text{t}(t)$ depends on time but $\Delta\dot{\nu}_\text{t}$ is a constant. The phenomenological glitch model above (or slight variants of it) have been established for a long time, having been used to model the first few glitches of the Crab pulsar \citep{boyntonetal1972} and the Vela pulsar \citep{downs1981}. Even now, it is still being used in pulsar timing softwares such as TEMPO2 \citep{edwardsHobbsManchester2006}.

Our analysis uses publicly available glitch data from the JBCA Glitch Catalogue \citep{espinozaetal2011}. Within this catalogue are values for $\frac{\Delta\nu(t_\text{g})}{\nu_0}$ and $\frac{\Delta\dot{\nu}(t_\text{g})}{\dot{\nu}_0}$ for each glitch of a given pulsar. Respectively, $\nu_0$ and $\dot{\nu}_0$ are the measured values of the spin frequency and the time derivative of the spin frequency immediately before the glitch. However, there is an issue using this data since we are not told exactly how much of $\Delta\nu(t_\text{g})$ and $\Delta\dot{\nu}(t_\text{g})$ is due to a transient part and how much of it is due to a permanent part. We therefore require more data and for us, it will be in the form of the healing parameter, $Q$.

By definition, $Q$ is defined as
\begin{equation} 
\label{def: Q}
Q \equiv \frac{\Delta\nu_\text{t} }{\Delta\nu(t_\text{g})}~,
\end{equation} 
where the numerator represents the change in spin frequency which will ``heal'' at $t \gg t_\text{g}$ and the denominator represents the total change in spin frequency at $t = t_\text{g}$ which is the sum of permanent and transient parts, as seen in Equation~(\ref{nuSum}). Therefore, a glitch which recovers completely has $Q=1$ and a glitch showing no recovery whatsoever has $Q=0$. 

We can substitute Equation~(\ref{def: Q}) into Equation~(\ref{def: delta dot nu t}) to relate our unknowns ($\Delta\dot{\nu}_\text{t}$, $\Delta\nu_\text{t}$) to our observables ($Q$, $\Delta\nu(t_\text{g})$) to get the relations
\begin{equation}
\label{unknownsToObservables}
\Delta\dot{\nu}_\text{t} \tau = - \Delta\nu_\text{t} = - Q \Delta\nu(t_\text{g})~.
\end{equation}

$\tau$ makes an appearance in Equation~(\ref{unknownsToObservables}) and it can either be treated as known or unknown. It is known when radio observations are frequent enough to acquire $\tau$ directly from the data. However, this is not the case for most glitches as the exponential recovery is sometimes missed, or if too few observations are made during the post-glitch recovery to allow a reliable fit. If it is unknown, then we can use Equation~(\ref{unknownsToObservables}) to approximate $\tau$, see Section~\ref{Discussion}. Throughout this analysis, we will treat $\tau$ as unknown but if it is known in reality, then our model provides an independent value of $\tau$ which can be checked for consistency and guide future transient CW searches.

\section{Ellipticity and gravitational wave strain}
\label{Maximum ellipticity and gravitational wave strain}

In this section, we focus on calculating the ellipticity and GW strain obtainable from transient mountains in terms of our observables. It will help at this point if we remind ourselves of the steps of the model: 1) a NS glitches, 2) a transient mountain immediately forms, 3) transient CWs are emitted whilst the transient mountain dissipates at a similar rate as the post-glitch recovery. We do not specify the cause of the glitch (e.g. superfluid unpinning \citep{andersonItoh1975}, starquakes \citep{ruderman1969}) and we do not attempt to explain the mechanism behind how the mountain forms or dissipates, but for a recent review on possible mechanisms for the formation of mountains, see \citet{sieniatskaBejger2019}.

It is well-known that a NS mountain will emit CWs at twice the NS's spin frequency, $f = 2\nu$ \citep[e.g.][]{shapiroTeukolsky1983}. A mountain also creates an extra braking torque on the system which we assume will explain the post-glitch recovery. Basic mechanics tells us that the torque, $\mathcal{N}$, is related to power (or in our case GW luminosity) by the equation $L = - \mathcal{N} \Omega$ (for $L>0$) where $\Omega$ is the angular velocity of the NS. It can also be shown \citep[e.g.][]{shapiroTeukolsky1983} that the GW luminosity due to a mountain is 
\begin{equation}
\label{def: L_GW}
L_\text{GW}(t) = \frac{(2\pi)^6}{10} \frac{G}{c^5} I^2 f^6 \varepsilon^2(t)~,
\end{equation}
where $I$ represents the moment of inertia about the rotation axis, $f$ is the GW frequency of the emitted CWs and $\varepsilon(t)$ is the dimensionless \textit{equatorial} ellipticity of the NS  (see Equation~(\ref{def: Equatorial ellipticity}) for the definition in terms of moment of inertias about the principal axes).

By taking out a factor of $\Omega = 2\pi \nu = \pi f$ from Equation~(\ref{def: L_GW}), multiplying throughout by $-1$ and using $f \approx 2\nu_0$, we get the torque due to a NS mountain
\begin{equation}
\label{def: N_mountain}
\mathcal{N}_\text{mountain}(t) = - \frac{32(2\pi)^5}{5} \frac{G}{c^5}~I^2\nu^5_0\varepsilon^2(t)~.
\end{equation} 

Also from mechanics, we know $\mathcal{N} = \dot{I}\Omega + I\dot{\Omega}$ in general. However, as we show in Appendix~\ref{Appendix - Torque}, during the post-glitch recovery we can associate the change in torque solely to a change in the spin-down rate caused by a transient mountain which means
\begin{equation}
\label{def: delta N}
\Delta\mathcal{N}(t) = 2\pi I \Delta\dot{\nu}_\text{t}(t)~,
\end{equation} 
where we have ignored the effect of $\Delta \dot{I}$, $\Delta \Omega$ and $\Delta I$ and specialised to transient mountains only. It is worth noting that the internal superfluid is important during glitches and can cause changes to the NS's moment of inertia \citep[e.g.][]{linkepsteinlattimer1999, anderssonetal2012}. However, our model only concerns the change of the shape of the NS during the post-glitch recovery and omits details about the interior, as the change in shape is what we ascribe to the post-glitch recovery. 


Since our model associates the change in torque purely to a transient mountain, we equate the left hand sides of Equations~(\ref{def: N_mountain}) and (\ref{def: delta N}). This allows us to find an expression for the ellipticity for $t \ge t_\text{g}$ which is
\begin{equation}
\label{epsilonSquared}
\varepsilon(t) = \sqrt{- \frac{5}{32(2\pi)^4} \frac{c^5}{G} \frac{1}{I} \frac{\Delta\dot{\nu}_\text{t}(t)}{\nu_0^{5}}} ~.
\end{equation} 

Note that this means our model will only apply to pulsars rotating fast enough, as slow rotators will give a value of $\varepsilon(t)$ which would be problematically large in terms of the physics. Since $\Delta\dot{\nu}_\text{t}(t)$ is in the expression for $\varepsilon(t)$ and $\Delta\dot{\nu}_\text{t}(t)$ is unknown, we will set $\Delta\dot{\nu}_\text{t}(t) = \Delta\dot{\nu}(t)$ to give us an approximation for the ellipticity from a transient mountain. This means we assume the entire change in the spin-down rate is purely transient, i.e. no permanent mountains are formed which would be associated with $\Delta\dot{\nu}_\text{p}$ within the framework of our model. As mentioned later in a footnote of Section~\ref{Discussion}, this is true for most glitches besides the largest glitches from the Crab where there appears to be a linear relationship between $\Delta\dot{\nu}_\text{p}$ and $\Delta\nu$ \citep{lyneetal2015}. We can then rearrange in terms of how the JBCA Glitch Catalogue reports glitch values, as well as setting $t = t_\text{g}$, to get the ellipticity at the moment of the glitch which is 
\begin{equation}
\label{epsilonSquaredMax}
\varepsilon_\text{approx}(t_\text{g}) = \sqrt{- \frac{5}{32(2\pi)^4} \frac{c^5}{G} \frac{1}{I} \frac{\dot{\nu}_0}{\nu_0^5} \left(\frac{\Delta\dot{\nu}(t_\text{g})}{\dot{\nu}_0} \right)}~.
\end{equation} 

Now that we have the ellipticity, we can calculate the GW strain, $h_0(t)$. In general, a given $\varepsilon(t)$ sources a GW strain of 
\begin{equation}
\label{def: h_0}
h_0(t) = (2\pi)^2 \frac{G}{c^4} \frac{I f^2}{d} \varepsilon(t)~,
\end{equation}
where $d$ is the distance to the source \citep[e.g.][]{jaranowskiKrolakSchutz1998}. We can then substitute the ellipticity, Equation~(\ref{epsilonSquared}), into Equation~(\ref{def: h_0}) to find the corresponding GW strain at $t \ge t_\text{g}$ which is
\begin{equation}
\label{h_0nudot}
h_{0}(t) = \sqrt{ - \frac{5}{2} \frac{G}{c^3} \frac{I}{d^2} \frac{\Delta\dot{\nu}_\text{t}}{\nu_0} } e^{-\frac{\Delta t}{2\tau}}~,
\end{equation} 
where we have used $\Delta\dot{\nu}_\text{t}(t) = \Delta\dot{\nu}_\text{t} e^{-\frac{\Delta t}{\tau}}$. Note that $h_0(t) \propto e^{-\frac{\Delta t}{2\tau}}$ and not $h_0(t) \propto e^{-\frac{\Delta t}{\tau}}$ as one might originally think. This implies $\tau_\text{GW} = 2\tau_\text{radio}$, see Section~\ref{Discussion}. This is due to $\Delta\dot{\nu}_\text{t}(t)$ being square-rooted in the expression for $\varepsilon(t)$, as per Equation~(\ref{epsilonSquared}). Finally, we can get an approximation for the GW strain at the moment of the glitch by using $\Delta\dot{\nu}_\text{t} = \Delta\dot{\nu}(t_\text{g})$ which gives
\begin{equation}
\label{h_0notAveraged}
h_{0, \text{approx}}(t_\text{g}) = \sqrt{ - \frac{5}{2} \frac{G}{c^3} \frac{I}{d^2} \frac{\dot{\nu}_0}{\nu_0} \left(\frac{\Delta\dot{\nu}(t_\text{g})}{\dot{\nu}_0} \right) }~ .
\end{equation}

\section{Total gravitational wave energy from the post-glitch recovery}
\label{Total gravitational wave energy from a glitch}

We now proceed on to calculating how much energy is available for the emission of GWs due to the loss of kinetic energy during the post-glitch recovery. We do not include in our calculation any energy which might be liberated during the glitch or required for the formation of transient mountains since these values are uncertain. We begin with the simple expression for the rotational kinetic energy, $E_\text{rot} = \frac{1}{2} I \Omega^2$. Any instantaneous loss (hence minus sign) of rotational kinetic energy we can write as a luminosity $L (>0)$ by differentiating with respect to time leading to
\begin{equation}
\label{def: L}
L = - 4 \pi^2 I \nu \dot{\nu}~.
\end{equation}

In particular, we are interested in the GW luminosity achievable due to a mountain which we attribute to having a negative $\Delta\dot{\nu}$. We can also get a change in the luminosity if there is a change in $I$ or $\nu$. However, as demonstrated in Appendices \ref{Appendix - Torque} and \ref{Appendix}, we are allowed to ignore the contributions due to $\Delta I$ and $\Delta \nu$ as these will be negligible.

The change in luminosity due to the glitch, $\Delta L$, can therefore be written as
\begin{equation}
\label{delta L}
\Delta L (t) = - 4 \pi^2 I \nu_0 \Delta \dot{\nu}_\text{t}(t)~,
\end{equation}
where we have put time-dependence into the equation and specialised to transient mountains only. This means we are only tracking the GWs due to transient mountains and not from any permanent mountains which arise due to a permanent change in the spin-down rate (these would give conventional CWs).

We can now integrate Equation (\ref{delta L}) across all time to get the GW energy due to a transient mountain and using $\Delta \dot{\nu}_\text{t}(t) = \Delta \dot{\nu}_\text{t}  e^{-\frac{\Delta t}{\tau}}$, we get
\begin{equation}
\label{E_GW using tau}
E_\text{GW} = - 4 \pi^2 I \nu_0 \Delta \dot{\nu}_\text{t} \tau = 4 \pi^2 I \nu_0 \Delta \nu_\text{t}~,
\end{equation}
and using Equation (\ref{unknownsToObservables}) we can relate the unknowns to our observables giving the final result of
\begin{equation}
\label{E_GW with Q and delta nu}
E_\text{GW} = 4 \pi^2 I \nu_0 Q \Delta \nu(t_\text{g})~,
\end{equation}
or when put in terms of how the JBCA Glitch Catalogue reports measurements, we get
\begin{equation}
\label{E_GW in observables}
E_\text{GW} = 4 \pi^2 I \nu_0^2 Q \left(\frac{\Delta \nu(t_\text{g})}{\nu_0}\right)~.
\end{equation}

Note that Equation~(\ref{E_GW with Q and delta nu}) can be seen as the answer you would get if you were to na{\"i}vely find the change in the rotational kinetic energy, but with an additional factor of $Q$ to account for the partial recovery of the glitch.

\section{Signal-to-noise ratio}
\label{Maximum signal-to-noise ratio}

Next, we relate the GW energy to the optimal\footnote{The SNR is optimal when the matched-filter exactly describes the data. Whenever there is a mismatch, the SNR becomes sub-optimal. See Fig.~1 of \citet{prixGiampanisMessenger2011}.} SNR achievable. We follow the steps of \citet{prixGiampanisMessenger2011} but apply it to our transient mountain model instead of their superfluid excess energy model. Additionally, we account for the effects of having multiple detectors and non-perpendicular interferometer arms which were not considered previously.

For a CW with a time-varying GW amplitude, $h_0(t)$, the geometrical-average of the optimal SNR squared, $\langle\rho_0^2\rangle$, is commonly written as
\begin{equation}
\label{def: averagedSNRsquared}
\langle\rho_0^2\rangle = \frac{4}{25} \frac{1}{S_\text{n}(f)} \int_{0}^{T_\text{obs}} h_0^2(t) dt~,
\end{equation}
where the angled brackets represents an average over geometrical parameters which are the right ascension, declination, polarisation angle and inclination of the source. This is what is responsible for the factor of $\frac{4}{25}$ which otherwise would not be there if the SNR was not averaged \citep{jaranowskiKrolakSchutz1998, prixGiampanisMessenger2011}. $S_\text{n}(f)$ is the GW detector's noise power spectral density which is a measure of how much noise the detector picks up at a given GW frequency, $f$. $T_\text{obs}$ is the duration for which we observe the source. We eventually set $T_\text{obs} \rightarrow \infty$ to ensure we capture the entire transient CW. 

Equation~(\ref{def: averagedSNRsquared}) is used frequently within the literature though it does not fully capture the effects which apply to us here in this analysis. The ET is a 3rd-generation GW detector which will consist of three identical interferometers, each with a pair of arms with an opening angle of 60\degree, arranged in a triangle such that two arms share the same side of the triangle \citep{freiseetal2009}. The fact that we have multiple detectors increases the SNR and an opening angle less than 90\degree~decreases the SNR. Combining both these effects changes Equation~(\ref{def: averagedSNRsquared}) into
\begin{equation}
\label{def: averagedSNRsquaredamended}
\langle\rho_0^2\rangle = \frac{4}{25} \frac{N \sin^2\zeta}{S_\text{n}(f)} \int_{0}^{T_\text{obs}} h_0^2(t) dt~,
\end{equation}
where $N$ is the number of independent interferometers and $\zeta$ is the opening angle between the arms of the interferometer. The $\sin^2\zeta$ factor can also be found in \citet{jaranowskiKrolakSchutz1998}. It is important we account for this as the $S_\text{n}(f)$ data for the ET is for a single 90\degree~interferometer\footnote{The ET-D sensitivity curve was taken from \url{http://www.et-gw.eu/index.php/etsensitivities}.}, even though in reality the ET is made up of three 60\degree~interferometers \citep{hildetal2011}.

We now need to evaluate the integral in Equation~(\ref{def: averagedSNRsquaredamended}). To do this, we directly substitute for $h_0(t)$ using Equation~(\ref{h_0nudot}) followed by Equation~(\ref{unknownsToObservables}) to get
\begin{equation}
\label{averagedSNRobservables}
\langle\rho_0^2\rangle= \frac{2}{5} \frac{G}{c^3} \frac{N \sin^2\zeta}{S_\text{n}(2\nu_0)} \frac{I}{d^2} Q \left(\frac{\Delta\nu(t_\text{g})}{\nu_0}\right)~,
\end{equation}
which is the SNR (squared) in terms of the observables within our model. 

Alternatively, we can utilise the GW energy we calculated in Section~\ref{Total gravitational wave energy from a glitch}. Equation~(\ref{def: L_GW}) gives the GW luminosity as a function of $\varepsilon(t)$, but $\varepsilon(t)$ is related to $h_0(t)$ through Equation~(\ref{def: h_0}). We can therefore write the GW luminosity as a function of $h_0(t)$ and integrate to get an expression for the GW energy. This gives
\begin{equation}
\label{E_GWintegral}
E_\text{GW} = \frac{2\pi^2}{5} \frac{c^3}{G} f^2 d^2 \int_{0}^{T_\text{obs}} h_0^2(t) dt~.
\end{equation}

We rearrange for the integral and substitute the integral into Equation~(\ref{def: averagedSNRsquaredamended}) to get
\begin{equation}
\label{averagedSNRinE_GW}
\langle\rho_0^2\rangle = \frac{2}{5\pi^2} \frac{G}{c^3} \frac{N \sin^2\zeta}{S_\text{n}(f)} \frac{E_\text{GW}}{f^2 d^2}~,
\end{equation}

Finally, we substitute in the GW energy calculated in Equation~(\ref{E_GW with Q and delta nu}) and we get the same answer as Equation~(\ref{averagedSNRobservables}). This alternative method is perhaps more powerful than the first as Equations~(\ref{E_GWintegral}) and (\ref{averagedSNRinE_GW}) are completely agnostic with regard to the time-dependence of $h_0(t)$ (since we substituted the entire integral). For the same reason, Equation~(\ref{averagedSNRinE_GW}) is valid even if we integrate over a short period of time. In other words, it means that we accumulate more SNR the longer the observation is. Equation~(\ref{averagedSNRinE_GW}) agrees with Equation~(4) in \citet{prixGiampanisMessenger2011} besides the factor of $N \sin^2\zeta$ which we added in here.

To bring familiarity to all this, we can also express our results in terms of a measure of the GW strain used in burst searches, since after all, transient CWs are essentially just a long burst. The measure used is called the root-sum-squared of the GW \textit{amplitude}, $h_{0,\text{rss}}$, which is defined as
\begin{equation}
h_{0,\text{rss}}^2 \equiv \int_{0}^{T_\text{obs}} h_0^2(t) dt~.
\end{equation}

Similarly, we can define the geometrically-averaged $h_{0,\text{rss}}^2$ as
\begin{equation}
\label{def: averagedh0}
\langle h_{0,\text{rss}}^2 \rangle \equiv \frac{4}{25} \int_{0}^{T_\text{obs}} h_0^2(t) dt = \frac{4}{25} h_0^2(t_\text{g}) \tau~,
\end{equation}
such that
\begin{equation}
\label{alternativeSNR}
\langle\rho_0^2\rangle \equiv N \sin^2\zeta \frac{\langle h_{0,\text{rss}}^2 \rangle}{S_\text{n}(f)} = \frac{4}{25} N \sin^2\zeta \frac{h_0^2(t_\text{g}) \tau}{S_\text{n}(f)}~,
\end{equation}
where the last equalities in Equations (\ref{def: averagedh0}) and (\ref{alternativeSNR}) come from using our transient mountain model and setting $T_\text{obs} \rightarrow \infty$. The first equality in Equation~(\ref{alternativeSNR}) can be seen from combining Equation~(\ref{def: averagedSNRsquaredamended}) and the definition in Equation~(\ref{def: averagedh0}).

We can check for consistency by substituting our expression for $h_0(t_\text{g})$ from Equation~(\ref{def: h_0}) and then using Equation~(\ref{unknownsToObservables}) to get the SNR in terms of observables. After doing that, we find that the right hand side of Equation~(\ref{alternativeSNR}) gives the same answer as Equation~(\ref{averagedSNRobservables}). In terms of our observables, $\langle h_{0,\text{rss}}^2 \rangle$ can explicitly be written as
\begin{equation}
\label{rssObservables}
\langle h_{0,\text{rss}}^2 \rangle = \frac{2}{5} \frac{G}{c^3} \frac{I}{d^2} Q \left(\frac{\Delta\nu(t_\text{g})}{\nu_0} \right)~.
\end{equation} 

The benefit of calculating $\langle h_{0,\text{rss}}^2 \rangle$ is that it has the same units as $S_\text{n}(f)$ which are units of $\text{Hz}^{-1}$. This then allows us to plot both quantities on the same set of axes allowing a visual comparison between the two terms. In fact, if we plot $\sqrt{\langle h_{0,\text{rss}}^2 \rangle}$ and $\sqrt{\frac{S_\text{n}(f)}{N \sin^2\zeta}}$ on the same axes, then the ratio of the two gives exactly the SNR. 

In reality, there is some SNR threshold, $\rho_\text{thres}$, which we must exceed before confidently accepting a detection. It varies depending on what search method is used, whether the data is stacked coherently and how large the search parameter space is \citep{walshetal2016, dreissigackerPrixWette2018}. For a targeted coherent CW search, this threshold\footnote{This particular SNR threshold gives a single trial false alarm rate of 1\% and a false dismissal rate of 10\%.} is $\rho_\text{thres} \approx 11.4$ \citep{abbottetal2004threshold}. Therefore, if we plot $\sqrt{\langle h_{0,\text{rss}}^2 \rangle}$ and $\sqrt{\frac{S_\text{n}(f) \rho_\text{thres}^2}{N \sin^2\zeta}}$ on the same axes, any signal which lies above the modified sensitivity curve will be confidently detectable, with a SNR greater than $11.4$.

\section{Applying the transient mountain model to data}
\label{Results}

We will now take our model and apply it to observed radio data. First we need to select which pulsars to use in our model. Obviously the pulsar must glitch so that constrains us to 190 pulsars\footnote{Value taken from \url{http://www.jb.man.ac.uk/pulsar/glitches/gTable.html} on 24th August 2020.}. However, the most limiting factor requires us to resolve the recovery of the glitch which requires a high cadence of observations, i.e.~pulsars which are observed frequently enough to see changes due to the glitch recovery. There are two outstanding candidates which satisfy these constraints and they are the Crab pulsar (B0531+21) and the Vela pulsar (B0833-45). These two pulsars are observed daily at the Jodrell Bank Observatory \citep{lyneetal2015} and at the Mount Pleasant Radio Observatory \citep{dodsonLewisMcCulloch2007} respectively. Therefore, our analysis will focus on the Crab and Vela pulsars, though the model is applicable to any glitching pulsar where $Q$ can be obtained and is rotating fast enough to prevent unphysically large mountains, see Equation~(\ref{epsilonSquared}).

We used radio data from three main sources: JBCA Glitch Catalogue for $\frac{\Delta\nu(t_\text{g})}{\nu_0}$ and $\frac{\Delta\dot{\nu}(t_\text{g})}{\dot{\nu}_0}$ \citep{espinozaetal2011}, \citet{crawforddemianski2003} for $Q$ and the ATNF Pulsar Database for $\nu_0$, $\dot{\nu}_0$ and $d$ \citep{manchesteretal2005}. The value for $\nu_0$ needed to be modified to account for the secular spin-down of the NS\footnote{We found that if one used the value of $\nu_0$ for the Crab from the ATNF Pulsar Catalogue \textit{without} accounting for secular spin-down, then the GW frequency fell within the 60~Hz mains power line spike in the sensitivity curves of the O2 detectors.}, but $\dot{\nu}_0$ and $d$ were taken as what was reported in the ATNF Pulsar Catalogue. The secular spin-down model we used included the first time derivative of the frequency only, namely
\begin{equation}
\label{secularspindown}
\nu_{0}(t_\text{g}) = \nu_{0,\text{ATNF}} + \dot{\nu}_{0,\text{ATNF}} \cdot (t_\text{g} - t_\text{ATNF})~,
\end{equation} 
where the subscript ``ATNF'' represents the value taken from the ATNF Pulsar Catalogue. $t_\text{ATNF}$ is the epoch corresponding to when the ATNF values were calculated. This was mid-1991 for the Crab and early-2000 for Vela \citep{manchesteretal2005}. The left hand side of Equation~(\ref{secularspindown}) is what was used for $\nu_0$ in all calculations. 

$t_\text{g}$ was set as the date of the glitch when calculating $E_\text{GW}$, $\varepsilon_\text{approx}(t_\text{g})$ and $h_{0,\text{approx}}(t_\text{g})$. However, when calculating SNRs for the different detectors, $t_\text{g} = \text{MJD}~57856$ for O2 detectors, \mbox{$t_\text{g} = \text{MJD}~59761$} for aLIGO at design sensitivity and \mbox{$t_\text{g} = \text{MJD}~64693$} for the ET. These correspond to the middle of the O2 run, the middle of 2022 and the middle of the 2030s respectively, which are the approximate times for when these detectors may be operational. This step is required since we have $S_\text{n}(2\nu_{0})$ in the denominator of the SNR and $\nu_0$ depends on the epoch you observe it.

The values for $\nu_0$, $\dot{\nu}_0$ and $d$ from the ATNF Pulsar Catalogue
are found in Table~\ref{tab:CrabVelaSpins} and the remaining data from the other two data sources can be found in Tables~\ref{tab:CrabResults} and \ref{tab:VelaResults}. Any unknown $Q$'s were set to the average $Q$ from existing known values. For the Crab, this was $Q \approx 0.84$ and for Vela, it was $Q \approx 0.17$.

\begin{table}
	\centering
	\caption{The values for the spin frequency, its time derivative and the distance to the Crab and Vela pulsars. These values were calculated in mid-1991 and early-2000 for the Crab and Vela pulsars respectively. Data taken from the ATNF Pulsar Database \citep{manchesteretal2005}.}
	\label{tab:CrabVelaSpins}
	\begin{tabular}{c c} 
		\hline \hline \\[-8pt]
		\textbf{Crab} & \textbf{Vela} \\
		\hline \\[-6pt]
		$\nu_0 = 29.947$~Hz & $\nu_0 = 11.195$~Hz \\
		$\dot{\nu}_0 = - 3.77 \times 10^{-10}$~Hz~s$^{-1}$ & $\dot{\nu}_0 = - 1.567 \times 10^{-11}$~Hz~s$^{-1}$ \\
		$d = 2.00$~kpc & $d = 0.28$~kpc \\ \\[-8pt]
		\hline \hline
	\end{tabular}
\end{table}

\citet{crawforddemianski2003} provides a comprehensive list of $Q$'s compiled from a large collection of literature. Each glitch has a $Q$ value and for some glitches, there are several $Q$'s, each from a different research group. Whenever there was more than one $Q$ for a single glitch, we used the average which is the value we report in Tables~\ref{tab:CrabResults} and \ref{tab:VelaResults}. 

As for the GW side of the analysis, we will look at the SNRs achievable in the Hanford and Livingston detectors (during the O2 observation run), aLIGO at its design sensitivity and the ET in its `D' configuration\footnote{The sensitivity curves of Hanford, Livingston and aLIGO at design sensitivity were taken from \url{https://dcc.ligo.org/LIGO-G1701570/public}, \mbox{\url{https://dcc.ligo.org/LIGO-G1701571/public}} and \url{https://dcc.ligo.org/LIGO-T1800044/public} respectively.}. Like the sensitivity curve of the ET, the aLIGO design sensitivity curve is for a single interferometer \citep{LSC2015} and since two 90\degree~interferometers make up aLIGO, we set $N_{\text{aLIGO}} = 2$ and $\zeta = 90\degree$. As established in Section~\ref{Maximum signal-to-noise ratio}, $N_{\text{ET}} = 3$ and $\zeta = 60\degree$ for the ET. We will treat each of the Hanford and Livingston detectors in O2 as individual interferometers since they have slightly different sensitivity curves, hence, $N = 1$ and $\zeta = 90\degree$ for both. One can simply multiply the SNR of either Hanford or Livingston by $\sqrt{2}$ to get an approximate combined SNR (or add in quadrature for a more accurate value), but as we will see, this makes little difference in detecting transient CWs outlined in our model for these two detectors.

Together, this data was used to calculate $E_\text{GW}$, $\sqrt{\langle \rho_0^2 \rangle}$, $\varepsilon_\text{approx}(t_\text{g})$, $h_{0,\text{approx}}(t_\text{g})$ and $\sqrt{\langle h_{0,\text{rss}}^2 \rangle}$. The results are shown in Tables~\ref{tab:CrabResults} and \ref{tab:VelaResults}.

\begin{table*}
	\centering
	\caption{Summary of results for the Crab pulsar. Columns 1 - 3 are taken from the JBCA Glitch Catalogue \citep{espinozaetal2011} and Column 4 is the average $Q$ per glitch from \citet{crawforddemianski2003}. Columns 5 - 10 are calculated from Equations (\ref{E_GW in observables}), (\ref{averagedSNRobservables}) using $\zeta = 90\degree$, (\ref{averagedSNRobservables}) using $\zeta = 60\degree$, (\ref{epsilonSquaredMax}), (\ref{h_0notAveraged}) and (\ref{rssObservables}) respectively. The 3 values for the SNR for aLIGO represent Hanford (in O2), Livingston (in O2) and aLIGO at design sensitivity from left to right. $^\text{O2}$This glitch occurred during the O2 run of aLIGO. $^\text{O3}$This glitch occurred during the O3 run of aLIGO. $^*$These values of $Q$ were not found in the literature so the average value of the observed $Q$'s was used. $^\text{\textdagger}$Calculated from data in Shaw et al. (in prep).}
	\label{tab:CrabResults}
	\begin{tabular}{c c c c c c >{\centering}p{0.92cm} c c c c} 
		\hline\hline \\[-10pt]
		\multicolumn{10}{c}{\small\textbf{Crab}} \\ [-0.5pt]
		\hline \\[-8pt]
		\footnotesize MJD & \footnotesize $\frac{\Delta\nu(t_\text{g})}{\nu_0}$ & \footnotesize $\frac{\Delta\dot{\nu}(t_\text{g})}{\dot{\nu}_0}$ & \footnotesize $Q$ & \footnotesize $E_\text{GW}$ [erg] & \footnotesize $\sqrt{\langle \rho_0^2 \rangle}_\text{aLIGO}$ & \footnotesize $\sqrt{\langle \rho_0^2 \rangle}_\text{ET}$ & \footnotesize $\varepsilon_\text{approx}(t_\text{g})$ & \footnotesize $h_{0,\text{approx}}(t_\text{g})$ & \footnotesize $\sqrt{\langle h_{0,\text{rss}}^2 \rangle}$ [$\text{Hz}^{-\frac{1}{2}}$] \\[3pt]
		
		$40491.8$ & $7.2 \times 10^{-9}$ & $4.4 \times 10^{-4}$ & $0.84$ & $2.2 \times 10^{41}$ & $1.1,1.5,3.4$ & $36$ & $1.6 \times 10^{-5}$ & $3.0 \times 10^{-26}$ & $1.3 \times 10^{-23}$ \\
		$41161.98$ & $1.9 \times 10^{-9}$ & $1.7 \times 10^{-4}$ & $0.92$ & $6.3 \times 10^{40}$ & $0.6,0.8,1.8$ & $20$ & $9.7 \times 10^{-6}$ & $1.9 \times 10^{-26}$ & $6.7 \times 10^{-24}$ \\
		$41250.32$ & $2.1 \times 10^{-9}$ & $1.1 \times 10^{-4}$ & $0.84$ & $6.3 \times 10^{40}$ & $0.6,0.8,1.8$ & $20$ & $7.8 \times 10^{-6}$ & $1.5 \times 10^{-26}$ & $6.8 \times 10^{-24}$ \\
		$42447.26$ & $3.57 \times 10^{-8}$ & $1.6 \times 10^{-3}$ & $0.81$ & $1.0 \times 10^{42}$ & $2.4,3.2,7.4$ & $80$ & $3.0 \times 10^{-5}$ & $5.7 \times 10^{-26}$ & $2.7 \times 10^{-23}$ \\
		$46663.69$ & $6 \times 10^{-9}$ & $5 \times 10^{-4}$ & $1.00$ & $2 \times 10^{41}$ & $1,2,3$ & $40$ & $2 \times 10^{-5}$ & $3 \times 10^{-26}$ & $1 \times 10^{-23}$ \\
		$47767.504$ & $8.10 \times 10^{-8}$ & $3.4 \times 10^{-3}$ & $0.89$ & $2.6 \times 10^{42}$ & $3.8,5.1,11.6$ & $130$ & $4.4 \times 10^{-5}$ & $8.3 \times 10^{-26}$ & $4.3 \times 10^{-23}$ \\
		$48945.6$ & $4.2 \times 10^{-9}$ & $3.2 \times 10^{-4}$ & $0.87$ & $1.3 \times 10^{41}$ & $0.8,1.2,2.6$ & $28$ & $1.4 \times 10^{-5}$ & $2.6 \times 10^{-26}$ & $9.7 \times 10^{-24}$ \\
		$50020.04$ & $2.1 \times 10^{-9}$ & $2 \times 10^{-4}$ & $0.80$ & $5.9 \times 10^{40}$ & $0.6,0.8,1.8$ & $19$ & $1 \times 10^{-5}$ & $2 \times 10^{-26}$ & $6.6 \times 10^{-24}$ \\
		$50260.031$ & $3.19 \times 10^{-8}$ & $1.73 \times 10^{-3}$ & $0.68$ & $7.7 \times 10^{41}$ & $2.1,2.8,6.4$ & $69$ & $3.2 \times 10^{-5}$ & $6.0 \times 10^{-26}$ & $2.4 \times 10^{-23}$ \\
		$50458.94$ & $6.1 \times 10^{-9}$ & $1.1 \times 10^{-3}$ & $0.87$ & $1.9 \times 10^{41}$ & $1.0,1.4,3.2$ & $34$ & $2.5 \times 10^{-5}$ & $4.8 \times 10^{-26}$ & $1.2 \times 10^{-23}$ \\
		$50489.7$ & $8 \times 10^{-10}$ & $-2 \times 10^{-4}$ & $\dots$ & $\dots$ & $\dots$ & $\dots$ & $\dots$ & $\dots$ & $\dots$ \\
		$50812.59$ & $6.2 \times 10^{-9}$ & $6.2 \times 10^{-4}$ & $0.92$ & $2.0 \times 10^{41}$ & $1.1,1.4,3.3$ & $35$ & $1.9 \times 10^{-5}$ & $3.6 \times 10^{-26}$ & $1.2 \times 10^{-23}$ \\
		$51452.02$ & $6.8 \times 10^{-9}$ & $7 \times 10^{-4}$ & $0.83$ & $2.0 \times 10^{41}$ & $1.0,1.4,3.3$ & $35$ & $2 \times 10^{-5}$ & $4 \times 10^{-26}$ & $1.2 \times 10^{-23}$ \\
		$51740.656$ & $2.51 \times 10^{-8}$ & $2.9 \times 10^{-3}$ & $0.80$ & $7.1 \times 10^{41}$ & $2.0,2.7,6.1$ & $66$ & $4.1 \times 10^{-5}$ & $7.7 \times 10^{-26}$ & $2.3 \times 10^{-23}$ \\
		$51804.75$ & $3.5 \times 10^{-9}$ & $5.3 \times 10^{-4}$ & $0.84^*$ & $1.0 \times 10^{41}$ & $0.8,1.0,2.4$ & $25$ & $1.8 \times 10^{-5}$ & $3.3 \times 10^{-26}$ & $8.7 \times 10^{-24}$ \\
		$52084.072$ & $2.26 \times 10^{-8}$ & $2.07 \times 10^{-3}$ & $0.84^*$ & $6.7 \times 10^{41}$ & $1.9,2.6,6.0$ & $64$ & $3.5 \times 10^{-5}$ & $6.5 \times 10^{-26}$ & $2.2 \times 10^{-23}$ \\
		$52146.758$ & $8.9 \times 10^{-9}$ & $5.7 \times 10^{-4}$ & $0.84^*$ & $2.6 \times 10^{41}$ & $1.2,1.6,3.7$ & $40$ & $1.8 \times 10^{-5}$ & $3.4 \times 10^{-26}$ & $1.4 \times 10^{-23}$ \\
		$52498.257$ & $3.4 \times 10^{-9}$ & $7.0 \times 10^{-4}$ & $0.84^*$ & $1.0 \times 10^{41}$ & $0.7,1.0,2.3$ & $25$ & $2.0 \times 10^{-5}$ & $3.8 \times 10^{-26}$ & $8.6 \times 10^{-24}$ \\
		$52587.2$ & $1.7 \times 10^{-9}$ & $5 \times 10^{-4}$ & $0.84^*$ & $5.0 \times 10^{40}$ & $0.5,0.7,1.6$ & $18$ & $2 \times 10^{-5}$ & $3 \times 10^{-26}$ & $6.1 \times 10^{-24}$ \\
		$53067.078$ & $2.14 \times 10^{-7}$ & $6.2 \times 10^{-3}$ & $0.84^*$ & $6.3 \times 10^{42}$ & $5.9,8.1,18.4$ & $200$ & $6.0 \times 10^{-5}$ & $1.1 \times 10^{-25}$ & $6.8 \times 10^{-23}$ \\
		$53254.109$ & $4.9 \times 10^{-9}$ & $2 \times 10^{-4}$ & $0.84^*$ & $1.4 \times 10^{41}$ & $0.9,1.2,2.8$ & $30$ & $1 \times 10^{-5}$ & $2 \times 10^{-26}$ & $1.0 \times 10^{-23}$ \\
		$53331.17$ & $2.8 \times 10^{-9}$ & $7 \times 10^{-4}$ & $0.84^*$ & $8.2 \times 10^{40}$ & $0.7,0.9,2.1$ & $23$ & $2 \times 10^{-5}$ & $4 \times 10^{-26}$ & $7.8 \times 10^{-24}$ \\
		$53970.19$ & $2.18 \times 10^{-8}$ & $3.1 \times 10^{-3}$ & $0.84^*$ & $6.4 \times 10^{41}$ & $1.9,2.6,5.9$ & $63$ & $4.3 \times 10^{-5}$ & $8.0 \times 10^{-26}$ & $2.2 \times 10^{-23}$ \\
		$54580.38$ & $4.7 \times 10^{-9}$ & $2 \times 10^{-4}$ & $0.84^*$ & $1.4 \times 10^{41}$ & $0.9,1.2,2.7$ & $29$ & $1 \times 10^{-5}$ & $2 \times 10^{-26}$ & $1.0 \times 10^{-23}$ \\
		$55875.5$ & $4.92 \times 10^{-8}$ & $\dots$ & $0.84^*$ & $1.4 \times 10^{42}$ & $2.8,3.9,8.8$ & $95$ & $\dots$ & $\dots$ & $3.3 \times 10^{-23}$ \\
		$57839.92^\text{O2}$ & $2.14 \times 10^{-9}$ & $2.7 \times 10^{-4}$ & $0.70^\text{\textdagger}$ & $5.2 \times 10^{40}$ & $0.5,0.7,1.7$ & $18$ & $1.3 \times 10^{-5}$ & $2.4 \times 10^{-26}$ & $6.2 \times 10^{-24}$ \\
		$58064.555$ & $5.1637 \times 10^{-7}$ & $6.969 \times 10^{-3}$ & $0.84^*$ & $1.5 \times 10^{43}$ & $9.2,12.5,28.6$ & $310$ & $6.5 \times 10^{-5}$ & $1.2 \times 10^{-25}$ & $1.1 \times 10^{-22}$ \\
		$58237.357$ & $4.08 \times 10^{-9}$ & $4.6 \times 10^{-4}$ & $0.84^*$ & $1.2 \times 10^{41}$ & $0.8,1.1,2.5$ & $27$ & $1.7 \times 10^{-5}$ & $3.1 \times 10^{-26}$ & $9.4 \times 10^{-24}$ \\
		$58470.939$ & $2.36 \times 10^{-9}$ & $3.60 \times 10^{-4}$ & $0.84^*$ & $6.9 \times 10^{40}$ & $0.6,0.8,1.9$ & $21$ & $1.5 \times 10^{-5}$ & $2.7 \times 10^{-26}$ & $7.2 \times 10^{-24}$ \\ 
		$58687.59^\text{O3}$ & $3.60 \times 10^{-8}$ & $\dots$ & $0.84^*$ & $1.1 \times 10^{42}$ & $2.4, 3.3, 7.5$ & $81$ & $\dots$ & $\dots$ & $2.8 \times 10^{-23}$ \\[2pt]
		
		\hline\hline
	\end{tabular}
\end{table*}

\begin{table*}
	\centering
	\caption{This contains the same results as Table~\ref{tab:CrabResults} but for the Vela pulsar. $^\text{O2}$This glitch occurred during the O2 run of aLIGO. $^1$The data for this glitch was taken from \citet{xuetal2019}. $^*$These values of $Q$ were not found in the literature so the average value of the observed $Q$'s was used.}
	\label{tab:VelaResults}
	\begin{tabular}{>{\centering}p{1.35cm} c c c c c >{\centering}p{0.92cm} c c c c} 
		\hline\hline \\[-10pt]
		\multicolumn{10}{c}{\small\textbf{Vela}} \\ [-0.5pt]
		\hline \\[-8pt]
		\footnotesize MJD & \footnotesize $\frac{\Delta\nu(t_\text{g})}{\nu_0}$ & \footnotesize $\frac{\Delta\dot{\nu}(t_\text{g})}{\dot{\nu}_0}$ & \footnotesize $Q$ & \footnotesize $E_\text{GW}$ [erg] & \footnotesize $\sqrt{\langle \rho_0^2 \rangle}_\text{aLIGO}$ & \footnotesize $\sqrt{\langle \rho_0^2 \rangle}_\text{ET}$ & \footnotesize $\varepsilon_\text{approx}(t_\text{g})$ & \footnotesize $h_{0,\text{approx}}(t_\text{g})$ & \footnotesize $\sqrt{\langle h_{0,\text{rss}}^2 \rangle}$ [$\text{Hz}^{-\frac{1}{2}}$] \\[3pt]
		
		$40280$ & $2.34 \times 10^{-6}$ & $1.0 \times 10^{-2}$ & $0.034$ & $3.9 \times 10^{41}$ & $2.1,6.1,24.8$ & $570$ & $1.8 \times 10^{-4}$ & $3.4 \times 10^{-25}$ & $3.2 \times 10^{-22}$ \\
		$41192$ & $2.05 \times 10^{-6}$ & $1.5 \times 10^{-2}$ & $0.035$ & $3.6 \times 10^{41}$ & $2.0,5.8,23.6$ & $540$ & $2.2 \times 10^{-4}$ & $4.2 \times 10^{-25}$ & $3.1 \times 10^{-22}$ \\
		$41312$ & $1.2 \times 10^{-8}$ & $3 \times 10^{-3}$ & $0.55$ & $3.3 \times 10^{40}$ & $0.6,1.8,7.1$ & $160$ & $1 \times 10^{-4}$ & $2 \times 10^{-25}$ & $9.4 \times 10^{-23}$ \\
		$42683$ & $1.99 \times 10^{-6}$ & $1.1 \times 10^{-2}$ & $0.21$ & $2.0 \times 10^{42}$ & $4.8,13.8,56.2$ & $1300$ & $1.9 \times 10^{-4}$ & $3.6 \times 10^{-25}$ & $7.4 \times 10^{-22}$ \\
		$43693$ & $3.06 \times 10^{-6}$ & $1.8 \times 10^{-2}$ & $0.12$ & $1.9 \times 10^{42}$ & $4.5,13.2,53.7$ & $1200$ & $2.4 \times 10^{-4}$ & $4.6 \times 10^{-25}$ & $7.0 \times 10^{-22}$ \\
		$44888.4$ & $1.145 \times 10^{-6}$ & $4.9 \times 10^{-2}$ & $0.177$ & $1.0 \times 10^{42}$ & $3.3,9.7,39.6$ & $910$ & $4.0 \times 10^{-4}$ & $7.5 \times 10^{-25}$ & $5.2 \times 10^{-22}$ \\
		$45192$ & $2.05 \times 10^{-6}$ & $2.3 \times 10^{-2}$ & $0.044$ & $4.5 \times 10^{41}$ & $2.2,6.5,26.4$ & $610$ & $2.7 \times 10^{-4}$ & $5.2 \times 10^{-25}$ & $3.5 \times 10^{-22}$ \\
		$46257.228$ & $1.601 \times 10^{-6}$ & $1.7 \times 10^{-2}$ & $0.158$ & $1.3 \times 10^{42}$ & $3.7,10.9,44.2$ & $1000$ & $2.3 \times 10^{-4}$ & $4.4 \times 10^{-25}$ & $5.8 \times 10^{-22}$ \\
		$47519.8036$ & $1.805 \times 10^{-6}$ & $7.7 \times 10^{-2}$ & $0.17^*$ & $1.5 \times 10^{42}$ & $4.1,12.0,48.7$ & $1100$ & $5.0 \times 10^{-4}$ & $9.5 \times 10^{-25}$ & $6.4 \times 10^{-22}$ \\
		$48457.382$ & $2.715 \times 10^{-6}$ & $6.0 \times 10^{-1}$ & $0.17^*$ & $2.3 \times 10^{42}$ & $5.1,14.7,59.7$ & $1400$ & $1.4 \times 10^{-3}$ & $2.6 \times 10^{-24}$ & $7.8 \times 10^{-22}$ \\
		$49559$ & $8.35 \times 10^{-7}$ & $0$ & $0.17^*$ & $7.0 \times 10^{41}$ & $2.8,8.2,33.1$ & $760$ & $\dots$ & $\dots$ & $4.3 \times 10^{-22}$ \\
		$49591.2$ & $1.99 \times 10^{-7}$ & $1.2 \times 10^{-1}$ & $0.17^*$ & $1.7 \times 10^{41}$ & $1.4,4.0,16.2$ & $370$ & $6.2 \times 10^{-4}$ & $1.2 \times 10^{-24}$ & $2.1 \times 10^{-22}$ \\
		$50369.345$ & $2.11 \times 10^{-6}$ & $5.95 \times 10^{-3}$ & $0.38$ & $4.0 \times 10^{42}$ & $6.7,19.4,78.7$ & $1800$ & $1.4 \times 10^{-4}$ & $2.6 \times 10^{-25}$ & $1.0 \times 10^{-21}$ \\
		$51559.319$ & $3.086 \times 10^{-6}$ & $6.736 \times 10^{-3}$ & $0.17^*$ & $2.6 \times 10^{42}$ & $5.4,15.7,63.7$ & $1500$ & $1.5 \times 10^{-4}$ & $2.8 \times 10^{-25}$ & $8.3 \times 10^{-22}$ \\
		$53193$ & $2.1 \times 10^{-6}$ & $\dots$ & $0.17^*$ & $1.8 \times 10^{42}$ & $4.4,12.9,52.5$ & $1200$ & $\dots$ & $\dots$ & $6.9 \times 10^{-22}$ \\
		$53960$ & $2.62 \times 10^{-6}$ & $2.3 \times 10^{-1}$ & $0.17^*$ & $2.2 \times 10^{42}$ & $5.0,14.4,58.7$ & $1300$ & $8.6 \times 10^{-4}$ & $1.6 \times 10^{-24}$ & $7.7 \times 10^{-22}$ \\
		$55408.8$ & $1.94 \times 10^{-6}$ & $7.5 \times 10^{-2}$ & $0.17^*$ & $1.6 \times 10^{42}$ & $4.3,12.4,50.5$ & $1200$ & $4.9 \times 10^{-4}$ & $9.3 \times 10^{-25}$ & $6.6 \times 10^{-22}$ \\
		$56556$ & $3.1 \times 10^{-6}$ & $1.48 \times 10^{-1}$ & $0.17^*$ & $2.6 \times 10^{42}$ & $5.4,15.7,63.8$ & $1500$ & $6.9 \times 10^{-4}$ & $1.3 \times 10^{-24}$ & $8.4 \times 10^{-22}$ \\
		$56922$ & $4 \times 10^{-10}$ & $1 \times 10^{-4}$ & $0.17^*$ & $3 \times 10^{38}$ & $0.1,0.2,0.7$ & $20$ & $2 \times 10^{-5}$ & $3 \times 10^{-26}$ & $1 \times 10^{-23}$ \\
		$57734.485^{\text{O2},1}$ & $1.431 \times 10^{-6}$ & $7.335 \times 10^{-2}$ & $0.0085$ & $6.0 \times 10^{40}$ & $0.8,2.4,9.7$ & $220$ & $4.9 \times 10^{-4}$ & $9.2 \times 10^{-25}$ & $1.3 \times 10^{-22}$ \\[2pt]
		
		\hline\hline
	\end{tabular}
\end{table*}

\begin{figure*}
	\includegraphics[width=\linewidth]{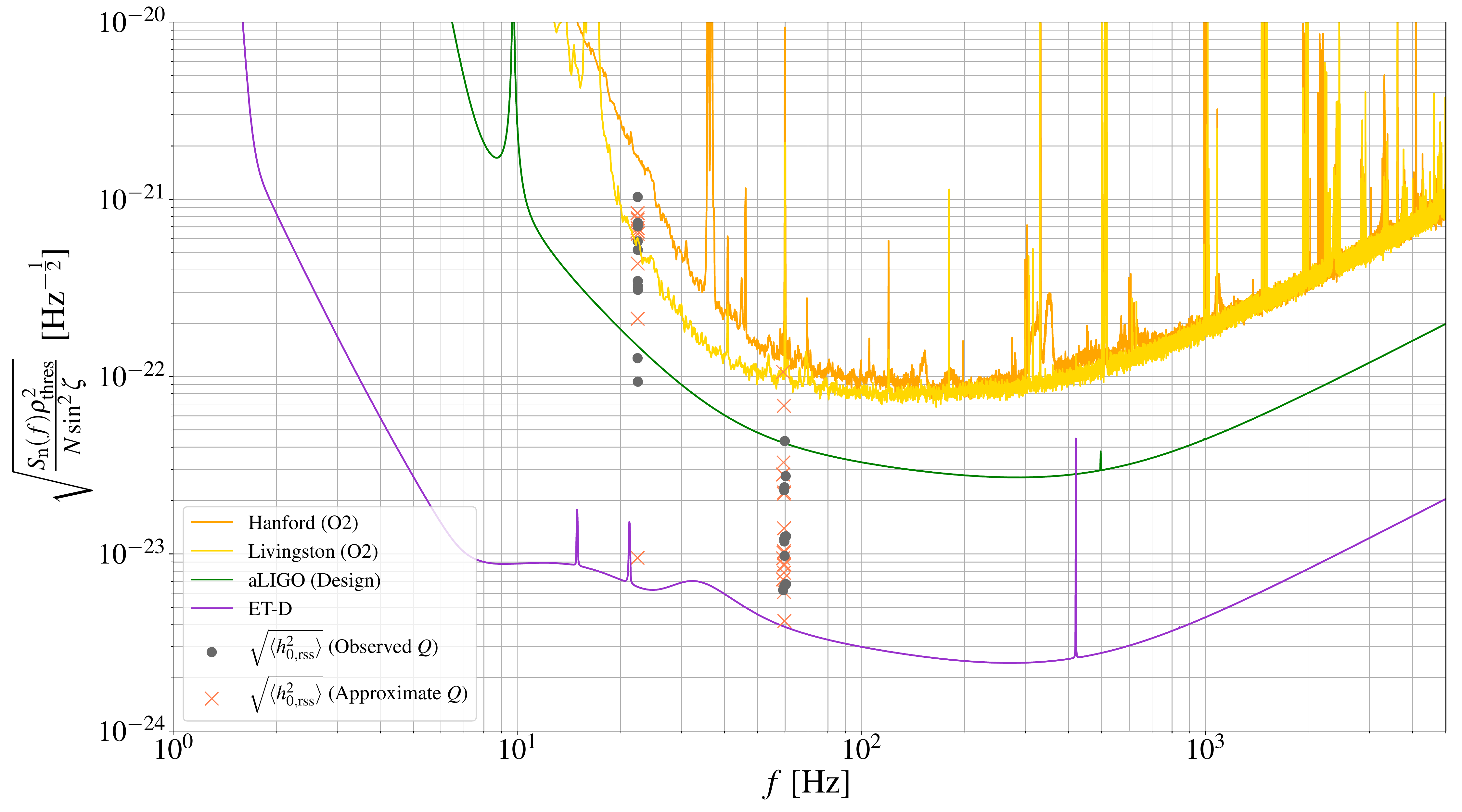}
	\caption{
		\label{fig:sensitivityCurvesAllDetectors}
		The above graph shows \textit{modified} sensitivity curves of ET in its `D' configuration (purple), aLIGO (green), Livingston in O2 (yellow) and Hanford in O2 (orange). The square root of the root-sum-squared GW amplitude of each transient CW is plotted as a grey filled circle. The light red crosses are the same but for the glitches we estimated $Q$ for. If a data-point lies above the modified sensitivity curve, then that transient CW signal would give a SNR greater than our threshold which is set to $\rho_\text{thres}= 11.4$, meaning it would be confidently detectable. The $x$-axis refers to the GW frequency which is twice the pulsar's spin frequency for a NS mountain.
	}
\end{figure*}

Finally, Fig.~\ref{fig:sensitivityCurvesAllDetectors} shows the detectability of the different glitch-induced transient mountains superimposed on modified sensitivity curves of the GW detectors of interest. The data-points (grey filled circles and light red crosses) represent the square root of the root-sum-squared GW amplitude, $\sqrt{\langle h_{0,\text{rss}}^2 \rangle}$, and the modified sensitivity curve is given by $\sqrt{\frac{S_\text{n}(f) \rho_\text{thres}^2}{N \sin^2\zeta}}$. The grey filled circles correspond to glitches we had $Q$ for, and the light red crosses refer to glitches where we did not have $Q$ and so the average $Q$ for that pulsar was used. If a data-point lies above the modified sensitivity curve, then the SNR would be greater than our threshold of 11.4 and would be classified as detectable. In reality, this detection threshold is only a guide and will differ depending on the confidence which you assign to the detection. The left bunch of data-points represents glitches from Vela and the right bunch belongs to the Crab. 

From this visual representation of the detectability, we can see immediately that the Hanford detector would not have detected transient CWs from glitch-induced transient mountains from the Crab and Vela pulsars, irrespective of whether the glitches occurred during the O2 run or not. To be clear, all but 2 glitches \textit{did not} happen in O2. This corresponds as expected with the numerical values of the SNR in Tables~\ref{tab:CrabResults}~and~\ref{tab:VelaResults}. 

For the other detectors, we see that some data-points are situated higher than the detector's sensitivity curve so we will look at the tabulated SNR values to identify these. For Livingston (O2), the Crab had 1 glitch and Vela had 10 glitches which exceeded the threshold SNR. These would have been detectable if they occurred in the O2 run, but again, those 11 glitches \textit{did not} occur in during the O2 run. 

There were only 2 glitches which \textit{did} occur during the O2 run, MJD~57839 (Crab) and MJD~57734 (Vela), and both had SNRs smaller than the 11.4 threshold in both the Hanford and Livingston detectors. This is consistent with the null findings of \citet{keiteletal2019}.

Furthermore, there was a glitch from the Crab (MJD 58687) which \textit{did} occur during the O3 observation run. Although the change in the spin-down rate due to the glitch has not yet been published, using our model and assuming the value of $Q \approx 0.84$, we predict transient CWs from this glitch should not be detectable, as \mbox{$\sqrt{\langle \rho_0^2 \rangle} = 7.5 < \rho_\text{thres}$} for the optimistic case of the O3 detectors having aLIGO's design sensitivity.


Moving onto aLIGO at design sensitivity, our calculations have shown that 3 Crab glitches would be detectable if these glitches were to occur when aLIGO is operational at design sensitivity. For Vela, 17 out of its 20 glitches would be detectable. This is indeed promising for upcoming transient CW searches. However, in general it seems like transient CWs from the Crab are unlikely to be detected, except for the very largest of glitches. Looking into the near future, we should focus on finding transient CWs from Vela's glitches, or any other sufficiently rapidly-rotating pulsar which is near, has a large glitch size~$\left(\frac{\Delta\nu(t_\text{g})}{\nu_0}\right)$, has a large glitch recovery ($Q$) or ideally some combination of all three.

Finally, for 3rd-generation detectors like the ET, we should almost certainly see transient CWs from both the Crab and Vela if glitch recoveries are even partly explained by our transient mountain model. If we do not see transient CWs with the next generation of GW detectors, one could put an upper limit on how much a glitch's recovery is due to a transient mountain.

%

\section{Discussion}
\label{Discussion}

It is interesting to point out that the total GW energy in our transient mountain model (Equation~(\ref{E_GW using tau})) depends on $\Delta\nu_\text{t}$, but $\Delta\nu_\text{t}$ can be written either as a product of $\Delta\nu(t_\text{g})$ and $Q$ or $\Delta\dot\nu_\text{t}$ and $\tau$, by Equation~(\ref{unknownsToObservables}). This highlights two different degeneracies within our model. The first is that the same GW energy, and hence SNR, can be obtained during the post-glitch recovery whether it be a large glitch with a small recovery, or a small glitch with a large recovery. This means that we should not neglect searching for transient CWs even if a pulsar has had a small glitch. There could very well be a significant transient CW signal, given the small glitch recovers by a large enough amount.

Secondly, there is a degeneracy between the transient change in the spin-down rate and the recovery time~scale. Since \mbox{$\varepsilon^2(t_\text{g}) \propto\Delta\dot\nu_\text{t}$} by Equation~(\ref{epsilonSquared}), the same SNR can be achieved with either a large transient mountain that decays away on a short time-scale, or a small transient mountain that decays on a long time-scale. Taken to the extreme, this statement is how conventional CW searches keep decreasing the upper limit on $\varepsilon$ or $h_0$ for a given pulsar, because not seeing a CW signal for a longer amount of time means the ellipticity must be smaller.

In previous sections, we briefly mentioned the idea of permanent mountains created at the moment of the glitch. They can exist from having a negative $\Delta\dot{\nu}_\text{p}$, but with the exception of a few large glitches of the Crab\footnote{It is well-known that large glitches of the Crab exhibit a permanent change to the spin-down rate \citep{lynePritchardGraham-Smith1993} but this is seemingly the only pulsar which shows this behaviour \citep{lyneShemarGraham-Smith2000}. The Crab had large glitches in the years 1975, 1989, 2011 and 2017 (MJD 42447.26, 47767.504, 55875.5 and 58064.555 respectively) with the first 3 showing a permanent change to $\dot{\nu}$ \citep{lyneetal2015}, and the last one too recent to confidently say whether there has been a permanent change to $\dot{\nu}$ \citep{shawetal2018}.}, many glitches do not normally show a non-zero $\Delta\dot{\nu}_\text{p}$. The only place where we needed to use $\Delta\dot{\nu}(t_\text{g})$ instead of $\Delta\dot{\nu}_\text{t}$ was in calculating $\varepsilon_\text{approx}(t_\text{g})$ and $h_{0,\text{approx}}(t_\text{g})$, where use of this approximation is clear from the subscript. If one had a value for $\Delta\dot{\nu}_\text{t}$, then using that to calculate $\varepsilon(t_\text{g})$ or $h_{0}(t_\text{g})$ would yield a more accurate result. The same can be said for including the effects of $\Delta\dot{\nu}_\text{p}$ in the total GW energy emitted. Permanent mountains would give off conventional CWs which contribute to a constant GW luminosity.

\citet{linkEpsteinBaym1992} argued that a negative $\Delta\dot{\nu}_\text{p}$ could not be explained with the vortex creep model\footnote{Alpar and collaborators soon added a "vortex depletion region" into their model to explain a negative $\Delta\dot{\nu}_\text{p}$ \citep{alparPines1993, alparetal1996}.} and concluded that there must be either a time-dependent torque during the glitch recovery \citep[e.g.][]{allenHorvath1997} or the glitch causes a change to the external torque. They preferred the idea of changing the external torque through a rearrangement of the magnetic field \citep{linkFrancoEpstein1998, francoLinkEpstein2000}, but our model, if one allows also for the formation of a permanent mountain, provides both the time-dependent torque from transient mountains and the (permanent) change to the external torque from permanent mountains. 


In our model, we can also make a prediction for the recovery time-scale by rearranging Equation~(\ref{unknownsToObservables}) and using the approximation $\Delta\dot{\nu}_\text{t} \approx \Delta\dot{\nu}(t_\text{g})$ to give
\begin{equation}
\label{def: t_approx}
\tau_\text{approx} \approx - \frac{Q \Delta\nu(t_\text{g})}{\Delta\dot{\nu}(t_\text{g})} = - \frac{\Delta\nu_\text{t}}{\Delta\dot{\nu}(t_\text{g})}~.
\end{equation}

If we were to include the effect of $\Delta\dot{\nu}_\text{p}$, then the recovery time-scale would be just a rearrangement of Equation~(\ref{def: delta dot nu t})
\begin{equation}
\label{def: tau}
\tau = - \frac{\Delta\nu_\text{t}}{\Delta\dot{\nu}_\text{t}} = - \frac{\Delta\nu_\text{t}}{Q'\Delta\dot{\nu}(t_\text{g})} = \frac{1}{Q'} \tau_\text{approx}~,
\end{equation}
where we follow \citet{weltevredeJohnstonEspinoza2011} in defining
\begin{equation}
\label{def: Q'}
Q' = \frac{\Delta\dot{\nu}_\text{t}}{\Delta\dot{\nu}(t_\text{g})} = \frac{\Delta\dot{\nu}_\text{t}}{\Delta\dot{\nu}_\text{t} + \Delta\dot{\nu}_\text{p}}~,
\end{equation}
which is analogous to $Q$ but for the spin-down rate. We can then look at $Q'$ as $\Delta\dot{\nu}_\text{p}$ varies, with $\Delta\dot{\nu}_\text{t}$ held as a negative constant. At $\Delta\dot{\nu}_\text{p} = 0$, we get $Q' = 1$ leading to $\tau = \tau_\text{approx}$ as expected. For the case of permanent mountains, $\Delta\dot{\nu}_\text{p} \lesssim 0$ so $Q' \lesssim 1$ and $\tau \gtrsim \tau_\text{approx}$. If there happens to be $\Delta\dot{\nu}_\text{p} \gtrsim 0$, then $Q' \gtrsim 1$ and $\tau \lesssim \tau_\text{approx}$. 

It is important to note that $\tau_\text{approx}$ is the approximate recovery time-scale of the post-glitch recovery which we see in \textit{radio}. As mentioned in Section~\ref{Maximum ellipticity and gravitational wave strain}, the GW recovery time-scale from a glitch, $\tau_\text{GW}$, defined by $h_0(t) \propto e^{-\frac{\Delta t}{\tau_\text{GW}}}$, is twice the recovery time-scale observed in the radio. Hence, for transient CW searches, one should use
\begin{equation}
\tau_\text{GW} = \frac{2}{Q'} \tau_\text{approx} = - \frac{2}{Q'} \frac{\nu_0}{\dot{\nu}_0} Q \left(\frac{\Delta\nu(t_\text{g})}{\nu_0}\right) \left(\frac{\Delta\dot{\nu}(t_\text{g})}{\dot{\nu}_0}\right)^{-1}~,
\end{equation}
or half of this if you are predicting the radio post-glitch recovery time-scale. Of course, if we already know $\tau$ from radio observations then $\tau_\text{GW}$ would simply be twice the value of $\tau$. This information should certainly help conduct transient CW searches and if, in the future, we detect transient CWs frequently, it could be that GW observations end up aiding radio astronomers in their discoveries instead of the other way around. Generally, GW searches are performed over some plausible range of damping times and not just at this single value \citep[e.g.][]{keiteletal2019}.

In both disciplines, it would be insightful to use typical values of $Q$ and $Q'$ when a pulsar glitches, so one could immediately predict how the GW waveform/radio timing residuals would appear in data under this model. For instance, we already calculated the average $Q$ for our two pulsars of interest, $Q_\text{Crab} \approx 0.84$ and $Q_\text{Vela} \approx 0.17$. We can try to do the same for $Q'$. Observationally, it appears that $\Delta\dot{\nu}_\text{p} < 0.2 \Delta\dot{\nu}(t_\text{g})$ for the Crab with the caveat that the glitch is not affected by the previous one \citep{espinozaetal2011, lyneetal2015}. This results in $0.8 < Q'_\text{Crab} < 1.0$ with larger Crab glitches taking values closer to the lower bound. It is harder to say for Vela as glitches tend to occur before the spin-down rate has had a chance to fully recover from the previous glitch \citep{lyneetal1996}. 


When we do calculate $\tau_\text{GW}$ using $Q' = 1$, we get average values of $\langle \tau_\text{GW, Crab} \rangle \approx 24$ days and $\langle \tau_\text{GW, Vela} \rangle \approx 298$ days. \citet{keiteletal2019} searched for transient CWs of up to 4 months in duration. One can see $\langle \tau_\text{GW, Vela} \rangle$ is longer than 4 months so if a typical glitch from Vela occurred, and it was uninterrupted by another glitch, then it would require a search longer than 4 months to track the first $e$-fold of its recovery. 

In a few cases, there were SNRs greater than our SNR threshold, $\rho_\text{thres}$. If these glitches occurred during a GW observation run (which they did not), then we could have had a detection if only a fraction of the total signal was detected. With an exponential recovery model, most of the GW energy is emitted within the first few $e$-folds of the signal, hence, we can reduce how long we need to integrate for. By using Equations~(\ref{def: averagedSNRsquaredamended}) and (\ref{h_0nudot}), we get 
\begin{equation}
\label{def: T_detect}
T_\text{detect} = - \tau \ln \left( 1 - \frac{\rho_\text{thres}^2}{\langle \rho_0^2 \rangle} \right)~,
\end{equation}
where $T_\text{detect}$ is the time it takes to confidently detect a signal and $\langle \rho_0^2 \rangle$ is given by Equation~(\ref{averagedSNRobservables}), which is the SNR (squared) you would get if you could capture the entire transient CW uninterrupted. $\langle \rho_0^2 \rangle$ can be predicted at the glitch if we use an approximate value of $Q$. One should note $\langle \rho_0^2 \rangle$ is different for every glitch, pulsar and GW detector so consequently $T_\text{detect}$ will be different in each case too. If a transient CW was detected confidently before $\Delta t = T_\text{detect}$ was reached, then our model would not be a viable explanation for that particular transient CW. In practical terms, if another glitch occurred before $\Delta t = T_\text{detect}$, but we were still able to confidently detect a transient CW signal from the first glitch, then this model fails to explain that transient CW. A reason for this to happen is if more GW energy is emitted immediately after the glitch and so an exponential form for the GW amplitude is not representative of the physics which cause the GWs.

As a proof of concept on how to use Equation~(\ref{def: T_detect}), we can set $T_\text{detect} = 4~\text{months}$ and $\tau$ to the average we would expect from Vela, so $\tau = \langle \tau \rangle = \frac{1}{2} \langle\tau_\text{GW, Vela}\rangle = 149~\text{days}$ to find what a 4 month search would be sensitive to if a typical Vela glitch with $\tau = 149~\text{days}$ were to occur. For $\rho_\text{thres} = 11.4$, the calculation shows that a 4 month search on a typical Vela glitch would only be detectable if the emitted transient CW had $\sqrt{ \langle \rho_0^2 \rangle} \ge 15.3$. Explicitly, this means a typical Vela glitch with $11.4 <\sqrt{ \langle \rho_0^2 \rangle} < 15.3$ would \textit{not} be detected with a 4 month search, though it would be if we could integrate for longer. Every glitch is different and so if $\tau$ was smaller, then the upper limit of the SNR range we are not sensitive to would reduce. It is clear that restricting the length of time searched over limits how many transient CWs we can detect. Therefore, appropriate modelling of transient CWs becomes important in selecting the upper temporal boundary for GW searches.

For the time it takes for a confident detection, the fraction of energy emitted is given by
\begin{equation}
\frac{E_\text{detect}}{E_\text{GW}} = \frac{\rho_\text{thres}^2}{\langle \rho_0^2 \rangle}~,
\end{equation}
where $E_\text{detect}$ is the accumulated GW energy emitted up to the detection time $T_\text{detect}$ and $E_\text{GW}$ is the total GW energy emitted if the glitch fully recovered. For the Vela case above, a confident detection within a 4 month observation corresponds to detecting at least 55\% of $E_\text{GW}$ for a Vela glitch. Any remaining kinetic energy lost during the post-glitch recovery could be attributed to elsewhere.

\begin{figure} 
	\centering
	\begin{subfigure}[b]{\columnwidth}
		\includegraphics[width=\columnwidth]{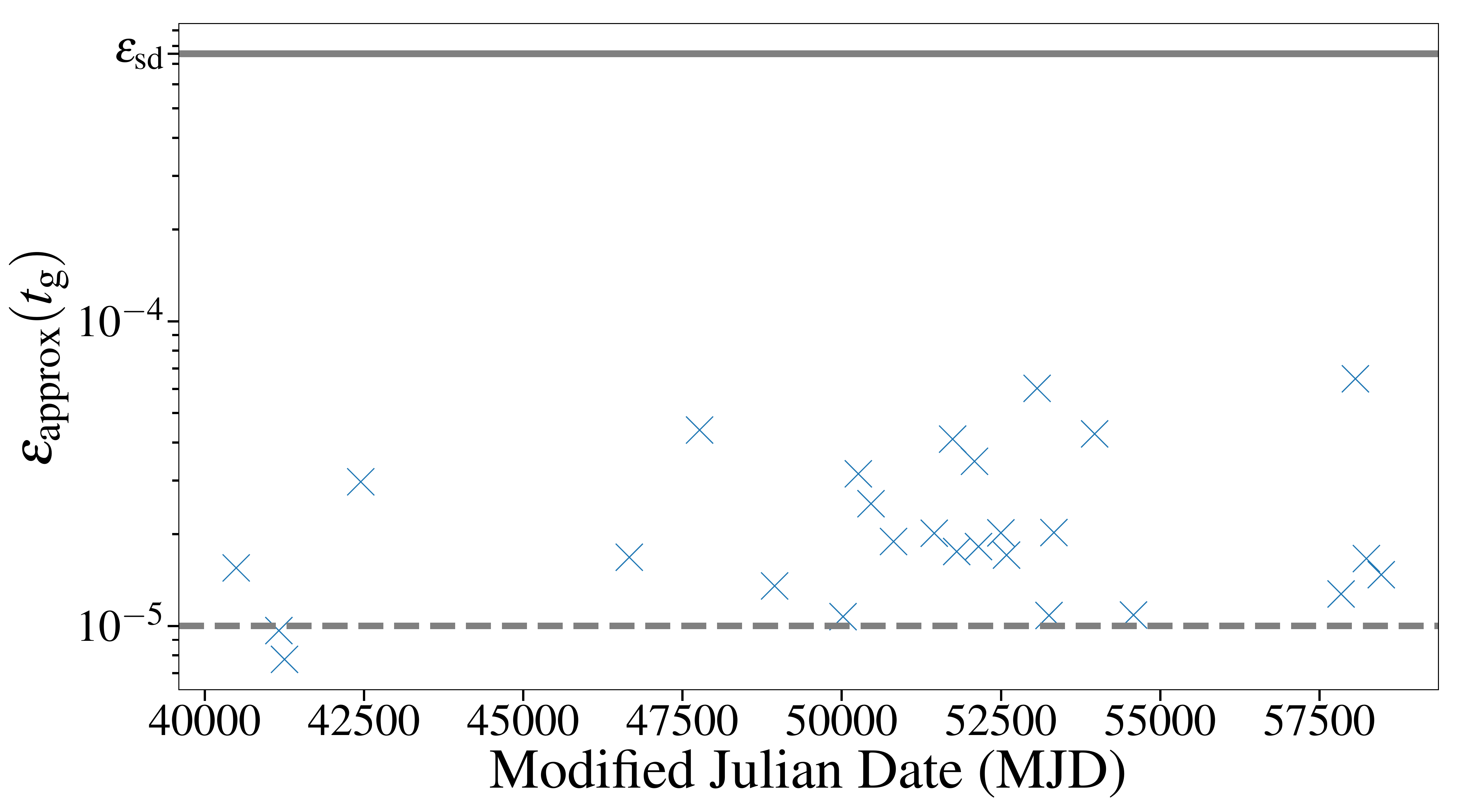}
	\end{subfigure} 
	\vspace{2pt}
	\begin{subfigure}[b]{\columnwidth}    
		\includegraphics[width=\columnwidth]{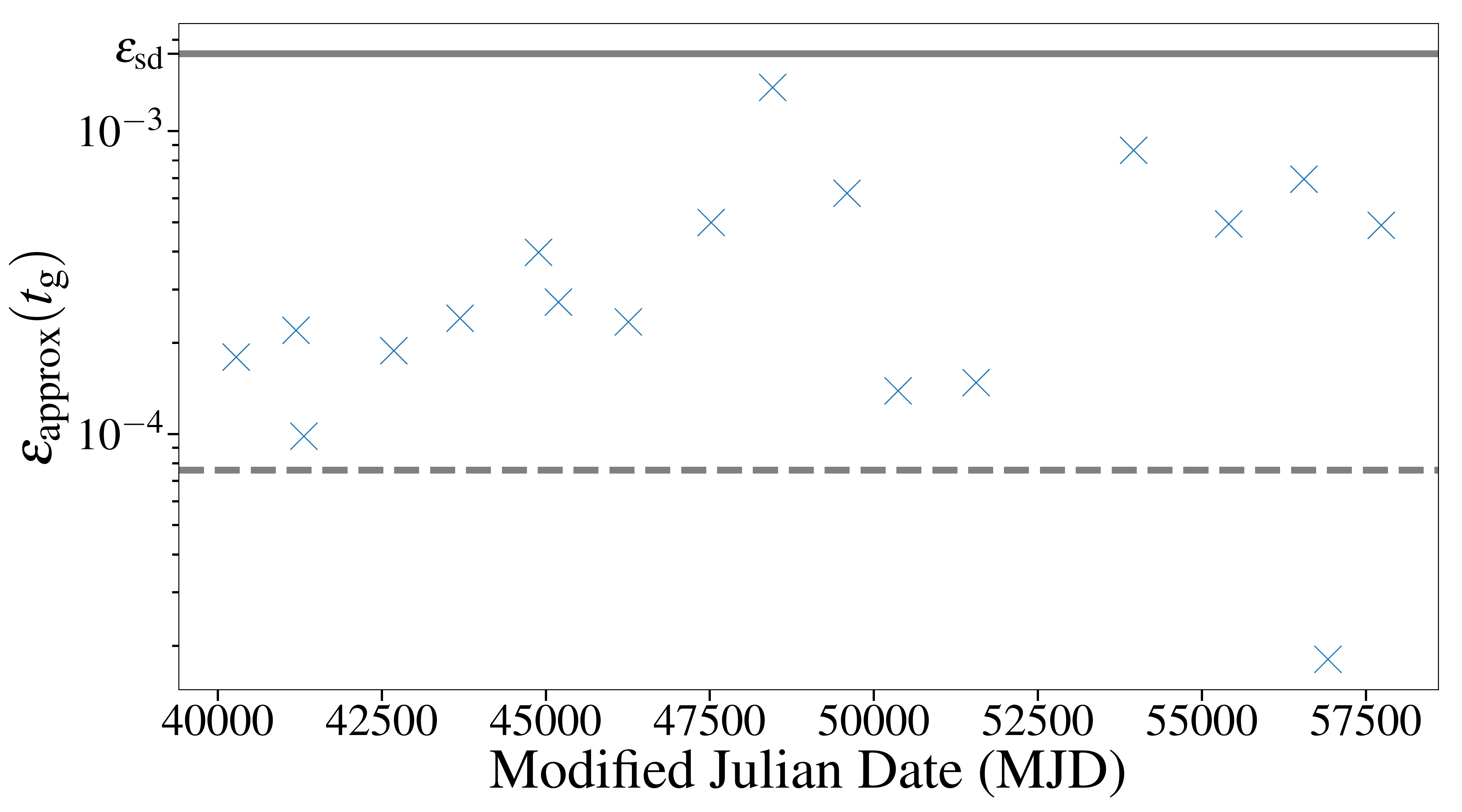}
	\end{subfigure} 
	\caption{
		\label{fig:epsilon}
		$\varepsilon_\text{approx}(t_\text{g})$ at various MJDs for glitches of the Crab (top) and Vela (bottom). The horizontal solid grey line represents the spin-down ellipticity required to explain the secular spin-down for that given pulsar. For the Crab, $\varepsilon_\text{sd} = 7.6 \times 10^{-4}$ and for Vela, $\varepsilon_\text{sd} = 1.8 \times 10^{-3}$. The horizontal dashed grey line represents the latest upper limit on the secular spin-down ellipticity as given in \citet{abbottetal2020upperlimits}. These upper limits are $\varepsilon_\text{upper} = 8.6 \times 10^{-6}$ and $\varepsilon_\text{upper} = 1.2 \times 10^{-4}$ for the Crab and Vela pulsars respectively.
	}
\end{figure}

For the ellipticities which we have calculated, it might be natural to want to see how they compare with the spin-down ellipticity of the pulsar. The spin-down ellipticity, $\varepsilon_\text{sd}$, is the ellipticity a pulsar would have if all of its secular spin-down were due to conventional CW emission. Mathematically, it is found from equating Equations~(\ref{def: L_GW}) and (\ref{def: L}) and solving for $\varepsilon(t) = \varepsilon_\text{sd}$, leading to $\varepsilon_\text{sd}^2 \propto \dot{\nu}_0$. Also, from Equation~(\ref{epsilonSquared}), we know $\varepsilon^2(t_\text{g}) \propto \Delta \dot{\nu}_\text{t}$. Both the constants of proportionality are the same and so we find the relation
\begin{equation}
\varepsilon^2(t_\text{g}) = \left(\frac{\Delta \dot{\nu}_\text{t}}{\dot{\nu}_0}\right) \varepsilon_\text{sd}^2~,
\end{equation}
or
\begin{equation}
\varepsilon^2_{\text{approx}}(t_\text{g}) = \left(\frac{\Delta \dot{\nu}(t_\text{g})}{\dot{\nu}_0}\right) \varepsilon_\text{sd}^2~,
\end{equation}
when we use the approximation $\Delta \dot{\nu}_\text{t} \approx \Delta \dot{\nu}(t_\text{g})$. Since $\left(\frac{\Delta \dot{\nu}(t_\text{g})}{\dot{\nu}_0}\right) < 1$, we can conclude $\varepsilon_{\text{approx}}(t_\text{g}) < \varepsilon_\text{sd}$ is always true. Therefore, the ellipticities of our transient mountains should never exceed the spin-down ellipticity. This is shown visually in Fig.~\ref{fig:epsilon} using data from the Crab and Vela. Also within Fig.~\ref{fig:epsilon} is a horizontal dashed line which represents the current upper limit on the spin-down ellipticity of the Crab and Vela pulsars, taken from \citet{abbottetal2020upperlimits}. It can be seen that many transient mountains have ellipticities which exceed this upper limit. However, this does not disprove our model, as the upper limit is calculated for a conventional CW with a duration on the order of years, whereas our shorter transient CWs may only have $\varepsilon > \varepsilon_\text{upper}$ for only some small fraction of time. There would not be enough signal emitted during this short time to accumulate enough SNR to be classified as a detection. This point just highlights the degeneracy between the size of the transient mountain and the time it takes to decay away. 

%

There is nothing in our model to suggest that we cannot have values of $Q>1$. This is in contrast to the vortex creep model where \citet{linkEpsteinBaym1992} showed only $Q<1$ was possible. Ultimately though, $Q<1$ is what is observed for most pulsars including the Crab and Vela pulsars. Looking at the definition of $Q$ in Equation~(\ref{def: Q}), $Q>1$ is mathematically equivalent to having $\Delta\nu_\text{p} < 0$ given \mbox{$\Delta\nu_\text{t}>0$}. This would appear as an ``over-recovery'' and has in fact been seen before in x-ray glitches of magnetars \citep{livingstoneKaspiGavriil2010, gavriilDibKaspi2011} and the high-magnetic-field radio pulsar J1119-6127 \citep{weltevredeJohnstonEspinoza2011, antonopoulouetlal2015}. However, it should be noted that magnetars generally rotate too slowly for our model to apply. Although a SNR could be calculated, any ellipticities would end up unphysically large. Additionally, J1119-6127 is one of the very few pulsars which have $\Delta\dot{\nu}_\text{p} > 0$. This feature cannot be explained by our model.

 For glitches with $Q>1$, the spin frequency after fully recovering is lower than the pre-glitch frequency. Therefore, if the recovery time-scale of such a glitch were to be shorter than the time between observations, then this over-recovery would be appear as an ``anti-glitch''. Such an anti-glitch was seen in \citet{archibaldetal2013}, but it was for a magnetar. One explanation for this was an enhanced electromagnetic outflow at the glitch, causing an increase on the external braking torque \citep[e.g.][]{tong2014}. If such an anti-glitch were to happen for a rapidly-rotating NS, then our model would associate the anti-glitch with a period of excess braking torque caused by the emission of GWs from a transient mountain. From Equation~(\ref{unknownsToObservables}), an arbitrarily large $Q$ can be achieved from having a transient mountain which is large enough, and/or lasts long enough. 

Recently, \citet{ashtonetal2019} looked at the rotational evolution of Vela during its most recent glitch in 2016 (MJD 57734). What was special about this glitch was that radio astronomers were observing the source at the moment it glitched, which had never been done before \citep{palfreymanetal2018}. In the frequency domain, \citet{ashtonetal2019} found a fast $\sim$$100~\text{s}$ recovery after the glitch and also a $\sim$$100~\text{s}$ spin-down prior to the glitch. To emphasize, these time-scales are much shorter than those considered in earlier sections. 

One interesting extension of our model would be to suggest the spin-down prior to the glitch is caused by the presence of a NS mountain. This mountain could either grow in size until some critical ellipticity which cracks the crust and triggers the glitch, or, it could be the instantaneous formation of a constant-sized (or growing) mountain which spins-down the normal component of the NS causing the lag between the normal and superfluid components to become critical, triggering vortex unpinning \citep{andersonItoh1975}. The post-glitch recovery would be associated with the mountain decaying away as previously established. Like before, we do not propose a formation mechanism for the mountain, rather, we only consider the consequences of such a mountain. Further analysis of $\ddot\nu$ would be required to distinguish if it was a growing mountain or a constant-sized mountain. 

This extension of the model implies that there would be GWs emitted at twice the NS's spin frequency \textit{prior} to the glitch, as well as after the glitch. If the crust-cracking idea is correct, it could also explain the symmetry of the situation since the crust would crack at the halfway point between it growing and decaying away, at the mountain's maximum size. Another symmetry is that the spin-down rate before and after the glitch appear to be the same magnitude. This would correspond to the mountain surviving the glitch event before decaying away. However, after putting in approximate values of the relevant observables into our model, we find that the ellipticity for this glitch would be around $\varepsilon = \mathcal{O}(10^{-1})$ at the time of the glitch which is large enough to crack the crust, but is too large for our model to be plausible. 

Unfortunately, Livingston was not operational at the moment of the glitch\footnote{\url{https://www.gw-openscience.org/detector_status/day/20161212/}} and after calculating the SNR for Hanford, the transient mountains would not have been detectable. Still, the shortest time-scale \citet{keiteletal2019} searched for was 0.5 days, which does not include the $\sim$$400~\text{s}$ GW duration this extended model would predict. Also, the all-sky search for 2 -- 500~s transients by \citet{abbottetal2019allsky} only looked at GW frequencies between 24 -- 2048~Hz, which does not include the $\sim$22~Hz GWs we would expect from Vela. It would certainly be interesting to see whether short transient CWs from the 2016 Vela glitch are still left undiscovered within O2 data.

Finally, our model is limited to glitches which have $\Delta\nu(t_\text{g}) > 0$ and $\Delta\dot{\nu}(t_\text{g}) < 0$. This is true for most glitches but there are a handful of other glitches which have $\Delta\nu(t_\text{g}) > 0$ and $\Delta\dot{\nu}(t_\text{g}) > 0$. So far, there is nothing in our model to explain such a glitch and would require more thought. 

\section{Conclusion}
\label{Conclusion}

To summarise, we have created a simple model whereby pulsar post-glitch recoveries seen in radio observations are attributed to the instantaneous formation of a transient mountain at the moment of the glitch. This mountain would create a braking torque on the NS spinning it back down, but not exclusively, to near pre-glitch spin frequencies via the emission of transient CWs. We calculated various quantities like the energy which could go into GW emission, the values for the SNR in different GW detectors as well as the ellipticity and GW strain which could be achieved from our model. All these quantities were expressed in terms of observable parameters from radio astronomy. We found that the greatest chance of detecting a transient CW required a large glitch size, a large glitch recovery and that the pulsar needed to be near to us in distance.

In general, we found that O2-era detectors would not have detected the majority of the Crab's or Vela's glitches, irrelevant of whether they actually occurred during the O2 run. There were however 2 glitches which did occur in O2 but both of these were not detectable according to our model. This is consistent with the recent findings of \cite{keiteletal2019}. Also, our model predicts that the Crab glitch which occurred in O3 will not be seen when the data from O3 is processed. One disappointing finding is that unless the Crab has a large glitch, it is unlikely we will see transient CWs from the Crab at aLIGO's design sensitivity. Fortunately though, most of Vela's glitches will be detectable at aLIGO's design sensitivity if our model is correct. Finally, with 3rd-generation detectors such as the ET, our results suggests we will see transient CWs from the glitches of both the Crab and Vela pulsars with the caveat that our model is correct, even if partially. If we do not see transient CWs from pulsar glitches at that point, then one could put an upper limit on how much transient mountains actually contribute to the post-glitch recovery. 



A consequence of our model is that large transient mountains which decay away quickly emit the same GW energy as smaller mountains which take longer to decay. This emphasises why it is important to search the uncharted territory between burst GWs and conventional CWs. To this end, we reported an estimate of the recovery time-scale for which transient CWs emitted via mountains would have. It was found that the GW recovery time-scale is exactly twice the post-glitch recovery time-scale found in radio observations, \mbox{i.e. $\tau_\text{GW} = 2 \tau_\text{radio}$}. We provided an expression for $\tau_\text{GW}$ dependent only on observables and could be quickly calculated either with or without the assumption of a permanent mountain being formed at the glitch. This will help guide future transient CW searches, not least, to give a theoretical explanation of a detection, if we are ever successful.

Finally, we provided a prediction for the minimum time required to accumulate enough SNR to warrant a detection, $T_\text{detect}$. If future searches do find transient CWs within a time shorter than $T_\text{detect}$, then this simple model, as it stands, will not be able to explain the newly-discovered phenomenon. 

\section*{Acknowledgements}

The authors would like to thank Gregory Ashton, David Keitel and Ben Stappers for fruitful discussions as well as Stefan Hild and Bangalore Sathyaprakash for useful discussions regarding the Einstein Telescope. We also thank Ben Shaw for kindly providing us with unpublished data for the Crab glitch which occurred on MJD~57839. GY acknowledges support from the EPSRC via grant number EP/N509747/1. DIJ acknowledges support from the STFC via grant number ST/R00045X/1. This research has made use of the JBCA Glitch Catalogue (\url{http://www.jb.man.ac.uk/pulsar/glitches.html}) and the ATNF Pulsar Catalogue (\url{https://www.atnf.csiro.au/research/pulsar/psrcat/}). This paper has been assigned document number LIGO-P2000222.

\section*{Data Availability}

All the data used in this paper is publicly available. Individual sources can be found in the main text and footnotes.




\bibliographystyle{mnras}
\bibliography{references} 




\appendix

\section{Simplifying the change in torque during the post-glitch recovery}
\label{Appendix - Torque}

This Appendix aims to show that we can simplify the general expression
for the change in torque on a NS due to a transient mountain
during the post-glitch recovery. We will exploit the size differences
between the fractional changes of several variables to allow us, to some
approximation, ignore all effects on the torque besides the change
in the spin-down rate.

Firstly, the torque $\mathcal{N}$ on the NS is given as
\begin{align}
\mathcal{N}(t) = \frac{d}{dt}(I(t)\Omega(t)) = \dot{I}(t)\Omega(t) + I(t) \dot{\Omega}(t)~,
\end{align} 
where $I$ is the moment of inertia about the rotation axis, $\Omega$
is the angular velocity of the NS and the dot represents a time
derivative. Then, the change in the torque due to a glitch is 
\begin{align}
\Delta \mathcal{N}(t) = 2\pi \left[ \nu_0 \Delta\dot{I}(t) + \dot{I}_0\Delta\nu(t) + \dot{\nu}_0\Delta I(t) + I_0\Delta\dot{\nu}(t) \right]~,
\end{align} 
for $t>t_\text{g}$ and where the subscript `0' represents the pre-glitch value. This is a general expression for the torque during the post-glitch recovery and each change of a variable is time-dependent. This means a change in $\dot{I}$, $\nu$, $I$ or $\dot{\nu}$ will have an effect on the torque. Our model associates the change in torque purely to a change in $\dot{\nu}$ so therefore we want to show that the first 3 terms are much smaller than the last in order for our assumption to be justified. The relevant ratios are
\begin{align}
\label{ratio1}
\frac{\text{Term 1}}{\text{Term 4}} &= \frac{\nu_0 \Delta\dot{I}(t)}{I_0\Delta\dot{\nu}(t)}~,  \\
\label{ratio2}
\frac{\text{Term 2}}{\text{Term 4}} &= \frac{\dot{I}_0\Delta\nu(t)}{I_0\Delta\dot{\nu}(t)}~,  \\
\label{ratio3}
\frac{\text{Term 3}}{\text{Term 4}} &= \frac{\dot{\nu}_0\Delta I(t)}{I_0\Delta\dot{\nu}(t)}~.
\end{align} 

These ratios are not trivial and so we will need explore further. In all ratios, the moment of inertia plays an important role so we will start with that. One can imagine that before a NS glitches, it has a stable, but still time-varying, moment of inertia, denoted by $I_\text{sec}(t)$ where the subscript is short for ``secular''. Then, immediately after the glitch it has a moment of inertia of $I(t_\text{g})$. The change of moment of inertia at the glitch is therefore $\Delta I(t_\text{g}) = I(t_\text{g}) - I_\text{sec}(t_\text{g}) = I(t_\text{g}) - I_0$. If the moment of inertia exponentially recovers at a rate $\tau$ (this can be verified by looking at Equations~(\ref{def: Meridional ellipticity}), (\ref{epsilonMeridional}) and (\ref{epsilonSquared}) but with the time-dependence kept in), then the moment of inertia of the NS goes as
\begin{align}
I(t) = I_\text{sec}(t) + \Delta I(t_\text{g})e^{-\frac{\Delta t}{\tau}} = I_\text{sec}(t) + \Delta I(t)~,
\end{align} 
for $t > t_\text{g}$, and when we differentiate with respect to time we get
\begin{align}
\dot{I}(t) = \dot{I}_\text{sec}(t) - \frac{\Delta I(t)}{\tau}~.
\end{align} 

We can therefore write 
\begin{align}
\Delta\dot{I}(t) \equiv \dot{I}(t) - \dot{I}_\text{sec}(t) = - \frac{\Delta I(t)}{\tau}~,
\end{align} 
which is the difference to the rate of change of the moment of inertia caused by the glitch. 

We also need to find an expression for $\dot{I}_0$ to allow us to simplify Equation~~(\ref{ratio2}). To do this, we need to look at the secular evolution of the NS's moment of inertia which is attributed to the slowing down of the NS over time. We can write the secular moment of inertia in terms of a small parameter, $\varepsilon_\Omega$, which parametrises how rotation causes a departure from the moment of inertia of a non-rotating NS. Mathematically, it is
\begin{align}
\label{def: Iepsilon}
I_\text{sec}(t) = I_\text{NR} (1+\varepsilon_\Omega(t))~,
\end{align} 
where $I_\text{NR}$ is the moment of inertia if the NS was not rotating. We immediately see that 
\begin{align}
\label{Idot}
\dot{I}_\text{sec}(t) = I_\text{NR} \dot{\varepsilon}_\Omega(t)~.
\end{align} 

In general, $\varepsilon_\Omega$ has the form
\begin{align}
\label{def: epsilonOmega}
\varepsilon_\Omega(t) = k\Omega^2(t)~,
\end{align} 
where $k$ is some constant of proportionality which depends on the detailed model of the NS. To get an idea of the size of $\varepsilon_\Omega$, we can take the simple case of a completely fluid and incompressible NS which \citet{baymPines1971} found has
\begin{align}
\varepsilon_\Omega(t) = \frac{5}{6} \frac{R^3}{GM} \Omega^2(t)~,
\end{align} 
which results in $\varepsilon_\Omega = 1.58 \times 10^{-4}$ for the Crab and $\varepsilon_\Omega = 2.21 \times 10^{-5}$ for Vela for values of the spin frequency as given in Table~\ref{tab:CrabVelaSpins} and using \mbox{$M = 1.4~\text{M}_\odot$} and $R = 10~\text{km}$.  Differentiating Equation~(\ref{def: epsilonOmega}) with respect to time and substituting into Equation~(\ref{Idot}), we find that
\begin{align}
\label{IdotNu}
\dot{I}_\text{sec}(t) = 2 I_\text{NR} \varepsilon_\Omega(t) \frac{\dot{\Omega}(t)}{\Omega(t)}~.
\end{align}

From Equation~(\ref{def: Iepsilon}), we can see that there is only a small correction between using $I_\text{NR}$ and say the moment of inertia just before the glitch, $I_0$ ($ =I_\text{sec}(t_\text{g})$), and so we can let $I_\text{NR} \approx I_0$. This then allows us to evaluate Equation~(\ref{IdotNu}) at the moment just before the glitch which gives
\begin{align}
\frac{\dot{I}_0}{I_0} = 2 \varepsilon_{\Omega,0} \frac{\dot{\nu}_0}{\nu_0}~.
\end{align}

We now have enough to re-write our ratios without the $\dot{I}_0$ and $\Delta\dot{I}$ terms, and we want all our ratios to be much less than~1. This results in 
\begin{align}
\label{inequality1}
\frac{\Delta I(t)}{I_0} &\ll - \frac{\tau\Delta\dot{\nu}(t)}{\nu_0}~,  \\
\label{inequality2}
2 \varepsilon_{\Omega,0} \frac{\Delta\nu(t)}{\nu_0} &\ll \frac{\Delta \dot{\nu}(t)}{\dot{\nu}_0}~, \\
\label{inequality3}
\frac{\Delta I(t)}{I_0} &\ll \frac{\Delta\dot{\nu}(t)}{\dot{\nu}_0}~,
\end{align} 
where each line represents the ratios in Equations~(\ref{ratio1}), (\ref{ratio2}) and (\ref{ratio3}) respectively. Note that these conditions are all still time-dependent. We can remove this by taking the conservative case of $\Delta I(t) = \Delta I(t_\text{g})$ and $\Delta \nu(t) = \Delta \nu(t_\text{g})$ as both these quantities are always smaller or equal to the value at the time of the glitch. On the right hand side, we can say that during the post-glitch recovery, $\Delta \dot{\nu}(t) \sim \Delta \dot{\nu}_\text{t}$. 
Then, using Equation~(\ref{unknownsToObservables}) in Equation~(\ref{inequality1}), and Equation~(\ref{def: Q'}) in Equations~(\ref{inequality2}) and (\ref{inequality3}), we finally get
\begin{align}
\label{finalinequality1}
\frac{\Delta I(t_\text{g})}{I_0} &\ll Q \left(\frac{\Delta\nu(t_\text{g})}{\nu_0}\right)~,  \\
\label{finalinequality2}
2 \varepsilon_{\Omega,0} \frac{\Delta\nu(t_\text{g})}{\nu_0} &\ll  Q' \left(\frac{\Delta \dot{\nu}(t_\text{g})}{\dot{\nu}_0} \right)~,  \\
\label{finalinequality3}
\frac{\Delta I(t_\text{g})}{I_0} &\ll Q' \left(\frac{\Delta\dot{\nu}(t_\text{g})}{\dot{\nu}_0}\right)~,
\end{align} 
where $Q$ is the healing parameter of the spin frequency and $Q'$ is the same but for the time derivative of the spin frequency. They are defined in Equations~(\ref{def: Q}) and (\ref{def: Q'}) respectively.


Typically, $Q$ and $Q'$ are on the order of unity and since the spin frequency of NSs vary by a small amount over secular time-scales, we can say $\varepsilon_{\Omega}$ varies by little too, meaning we are able to use the values of $\varepsilon_{\Omega}$ which we calculated earlier for $\varepsilon_{\Omega,0}$.


From Appendix~\ref{Appendix}, it was found empirically that \mbox{$\frac{\Delta I(t_\text{g})}{I_0} \ll \frac{\Delta\nu(t_\text{g})}{\nu_0}$} for both the Crab and Vela. This means Equation~(\ref{finalinequality1}) is satisfied. Then, looking at the JBCA Glitch Catalogue, we find that $\frac{\Delta\nu(t_\text{g})}{\nu_0}$ is about 4 - 6 orders of magnitude smaller than $\frac{\Delta\dot{\nu}(t_\text{g})}{\dot{\nu}_0}$ for both pulsars. Along with $\varepsilon_{\Omega,0} \sim 10^{-5}$ -- $10^{-4}$, Equation~(\ref{finalinequality2}) is comfortably satisfied. Finally, by simple logic it must be that $\frac{\Delta I(t_\text{g})}{I_0} \ll \frac{\Delta\dot{\nu}(t_\text{g})}{\dot{\nu}_0}$ so Equation~(\ref{finalinequality3}) is also satisfied. 

To summarise, we have shown Equations~(\ref{finalinequality1}) -- (\ref{finalinequality3}) to be true. This allows us to ignore the small effects of $\Delta\dot{I}$, $\Delta\nu$ and $\Delta I$ on the torque. As a result, the main contributor to the change in torque for our transient mountain model is due to a change in the spin-down rate, $\Delta\dot{\nu}$.

\section{The change in the moment of inertia due to the formation of a mountain}
\label{Appendix}

This Appendix addresses the question on whether the change in the moment of inertia about the rotation axis, due to the formation of a transient mountain, can be ignored when: 1) a NS rapidly spins-up during a glitch, and 2) when the NS recovers from a glitch. For a uniformly-dense and incompressible NS, the sudden formation of a mountain causes an increase to the moment of inertia about the rotation axis. This has the effect of decreasing the NS's spin frequency and the subsequent dissipation of the mountain during the recovery causes the NS to spin faster. We show empirically that these effects due to changes in the moment of inertia are negligible compared to the frequency changes caused intrinsically by glitches for the Crab and Vela pulsars. We use the results from this Appendix to simplify the calculation of the GW luminosity and torque during the post-glitch recovery.

For the sake of simplicity, we will assume the NS is of uniform density and is incompressible. This allows for simple analytic results to be calculated. Under the model described in Section~\ref{Introduction}, when a NS glitches, the NS forms a transient mountain turning it into a tri-axial (ellipsoidal) NS. This tri-axial NS will have a volume of 
\begin{align}
V = \frac{4}{3} \pi a_1 a_2 a_3~,
\end{align} 
where $a_1$, $a_2$ and  $a_3$ are the semi-major axis lengths along the $x$, $y$ and $z$ axes respectively, with the $z$-axis defined such that it is aligned with the rotation axis. The initial pre-glitch configuration, shown by a subscript `0', is generally axisymmetric with an initial volume of
\begin{align}
V_{0} = \frac{4}{3} \pi a_{1,0}^2a_{3,0}~,
\end{align} 
where we have used $a_{1,0} = a_{2,0}$ for an axisymmetric NS. Keeping all generality, the perturbed semi-major axis length can be written as the sum of the unperturbed semi-major axis length, $a_{i,0}$, and a small perturbation, $\delta a_i$, giving
\begin{align}
\label{def: a_i}
a_i = a_{i,0} + \delta a_i~,
\end{align} 
for $i = 1, 2, 3$. We are interested in spherical harmonic perturbations of the form $l = 2$, $m = 2$ which have the property $\delta a_2 = - \delta a_1$. When enforced with volume conservation, we find a 1-parameter family of solutions, with $\delta a_3$ easily calculable if required.

Let the moment of inertia about the rotation axis be $I_{zz}$. We have similar expressions along the $x$ and $y$ axes which lie along the remaining axes of symmetry. For a uniformly-dense ellipsoid of density $\bar{\rho}$, the moment of inertia about each of the axes can be written analytically as
\begin{align}
\label{def: Ellipsoidal MoI}
I_{xx} &= \frac{1}{5} M (a_2^2 + a_3^2)~, \nonumber \\ 
I_{yy} &= \frac{1}{5} M (a_1^2 + a_3^2)~, \\
I_{zz} &= \frac{1}{5} M (a_1^2 + a_2^2)~, \nonumber
\end{align} 
where $M = \bar{\rho} V$. Here, when we talk about NS mountains we refer to a non-axisymmetric deformation which leads to an \textit{equatorial} ellipticity, $\varepsilon_{\text{eq}}$. It is defined as
\begin{align}
\label{def: Equatorial ellipticity}
\varepsilon_{\text{eq}} \equiv \frac{I_{xx}-I_{yy}}{I_{zz}}~,
\end{align} 
which is a small dimensionless number parametrising how much the NS differs in length between its $x$ and $y$ axes within the equatorial plane (for a NS with a non-zero $I_{zz}$). 

There is another dimensionless parameter we can talk about which is the oblateness parameter, $\varepsilon_{\text{ob}}$, defined~as
\begin{align}
\label{def: Meridional ellipticity}
\varepsilon_{\text{ob}} \equiv \frac{I_{zz}-I_{zz,0}}{I_{zz,0}}~,
\end{align} 
where $I_{zz,0} = \frac{2}{5} M a_{1,0}^2$. This is, again, a small parameter but this time it measures the change in oblateness due to the sudden formation of a mountain. Oblateness can be seen as how elliptical the NS is in a plane perpendicular to the equatorial plane, i.e. within a meridional plane. 

We want to show that we can ignore the effects of the change in the moment of inertia about the $z$-axis whilst still obeying the conservation of angular momentum at the moment of the glitch. The change in the angular momentum at the moment of the glitch, $\Delta J (t_\text{g})$, is the sum of
\begin{align}
\Delta J (t_\text{g}) = 2 \pi \Delta I_{zz}(t_\text{g}) \nu_0 +  2 \pi I_{zz,0} \Delta\nu (t_\text{g})~,
\end{align} 
where the parentheses show the quantity is evaluated at the time of the glitch, immediately after the transient mountain has been created. If we can show the first term is smaller than the second, then we can conclude that the change in the moment of inertia can be ignored at the glitch without considerably affecting the ``usual'' change in angular momentum due to a change in the spin frequency. In other words, we want to show
\begin{align}
\label{MoI Condition}
\frac{\Delta I_{zz}(t_\text{g}) }{I_{zz,0}} \ll \frac{\Delta \nu (t_\text{g})}{\nu_0}~,
\end{align} 
where $\Delta I_{zz}(t_\text{g}) = I_{zz}(t_\text{g}) - I_{zz,0}$. The left hand side of Equation~(\ref{MoI Condition}) is the same as $\varepsilon_{\text{ob}}(t_\text{g})$ and the right hand side we know from the JBCA Glitch Catalogue. However, we do not immediately have a numerical value for $\varepsilon_{\text{ob}}(t_\text{g})$ but what we can get is a value for $\varepsilon_{\text{eq}}(t_\text{g})$. $\varepsilon_{\text{eq}}(t_\text{g})$ can be approximated using Equation~(\ref{epsilonSquaredMax}) from Section~\ref{Maximum ellipticity and gravitational wave strain}. Therefore, we proceed by relating $\varepsilon_{\text{ob}}$ to $\varepsilon_{\text{eq}}$ which would give us a numerical value to the left hand side of Equation~(\ref{MoI Condition}). 

After substituting Equations~(\ref{def: Ellipsoidal MoI})~and~(\ref{def: a_i}) into Equation~(\ref{def: Meridional ellipticity}), we get
\begin{align}
\label{Meridional ellipticity as function of da1 and da2}
\varepsilon_{\text{ob}} = \frac{- \left(a_{1,0}^2 - a_{2,0}^2\right) + \left( 2 a_{1,0} \delta a_1 + 2 a_{2,0} \delta a_2 \right) + \left( \delta a_1^2 + \delta a_2^2 \right)}{2a_{1,0}^2}~,
\end{align} 
and doing the same substitutions but for Equation~(\ref{def: Equatorial ellipticity}) gives
\begin{align}
\label{Equatorial ellipticity as function of da1 and da2}
\varepsilon_{\text{eq}} = \frac{\left(a_{2,0}^2 - a_{1,0}^2\right) + 2\left(a_{2,0} \delta a_2 - a_{1,0} \delta a_1 \right) + \left( \delta a_2^2 - \delta a_1^2 \right)}{\left( a_{1,0}^2 + a_{2,0}^2 \right) + 2\left(a_{1,0} \delta a_1 + a_{2,0} \delta a_2 \right) + \left(\delta a_1^2 + \delta a_2^2\right)}~.
\end{align} 

Both Equations~(\ref{Meridional ellipticity as function of da1 and da2}) and (\ref{Equatorial ellipticity as function of da1 and da2}) are general results. Then, we specialise to $\delta a_2 = - \delta a_1$ perturbations and account for the axisymmetric initial configuration, $a_{1,0} = a_{2,0}$, so that Equation~(\ref{Meridional ellipticity as function of da1 and da2}) becomes
\begin{align}
\varepsilon_{\text{ob}} = \left(\frac{\delta a_1}{a_{1,0}}\right)^2~,
\end{align} 
which is an exact result, and (the square of) Equation~(\ref{Equatorial ellipticity as function of da1 and da2}) becomes
\begin{align}
\varepsilon_{\text{eq}}^2 = 4\left(\frac{\delta a_1}{a_{1,0}}\right)^2 + \mathcal{O}\left[\left(\frac{\delta a_1}{a_{1,0}}\right)^4\right]~,
\end{align} 
where the $\mathcal{O}$ notation represents higher order terms we can ignore, coming from a Taylor expansion in $ \left(\frac{\delta a_1}{a_{1,0}}\right)$. Therefore to a very good approximation, we can say that
\begin{align}
\label{epsilonMeridional}
\varepsilon_{\text{ob}} \approx \frac{1}{4}\varepsilon_{\text{eq}}^2~,
\end{align} 
at all times during and after the glitch. We can now numerically evaluate Equation~(\ref{MoI Condition}) to see if it is satisfied. We take \mbox{$\varepsilon_{\text{eq}}(t_\text{g}) = \varepsilon_{\text{approx}}(t_\text{g})$} from Tables \ref{tab:CrabResults} and \ref{tab:VelaResults} to calculate $\varepsilon_{\text{ob}}(t_\text{g})$ and we compare $\varepsilon_{\text{ob}}(t_\text{g})$ as a percentage of $\frac{\Delta \nu (t_\text{g})}{\nu_0}$. The results are shown in Table~\ref{tab:Percentage MoI contribution}.

\begin{table*}
	\caption{This table has values of $\varepsilon_\text{ob}(t_\text{g})$ calculated from Equation~(\ref{epsilonMeridional}), the glitch sizes from the JCBA Glitch Catalogue \citep{espinozaetal2011} and the final two columns have the fraction of these two quantities expressed as a percentage, one without being divided by $Q$ and the other one with the division by $Q$. The 4th column represents the contribution that the change in the moment of inertia has on the angular momentum at moment of the glitch and 5th column represents the same but for the post-glitch recovery. $^\text{O2}$This glitch happened during the O2 run of aLIGO. $^\text{O3}$This glitch occurred during the O3 run of aLIGO. $^1$The data for this glitch was taken from \citet{xuetal2019}.}
	\label{tab:Percentage MoI contribution}
	\begin{tabular}{c c} 
		\begin{tabular}[t]{>{\centering}p{1.2cm} >{\centering}p{1.23cm} >{\centering}p{1.38cm} >{\centering}p{1.1cm} c} 
			\hline\hline \\[-8pt]
			\multicolumn{5}{c}{\small\textbf{Crab}} \\
			\hline \\[-6pt]
			\footnotesize MJD & \footnotesize $\varepsilon_{\text{ob}}(t_\text{g})$ & \footnotesize $\frac{\Delta\nu(t_\text{g})}{\nu_0}$ & \footnotesize $\frac{\varepsilon_{\text{ob}}(t_\text{g})\nu_0}{\Delta\nu(t_\text{g})}$[\%] & \footnotesize $\frac{\varepsilon_{\text{ob}}(t_\text{g})\nu_0}{Q\Delta\nu(t_\text{g})}$[\%] \\[4pt]
			
			$40491.8$ & $6.3 \times 10^{-11}$ & $7.2 \times 10^{-9}$ & $0.87$ & 1.0 \\
			$41161.98$ & $2.4 \times 10^{-11}$ & $1.9 \times 10^{-9}$ & $1.3$ & 1.4 \\
			$41250.32$ & $1.6 \times 10^{-11}$ & $2.1 \times 10^{-9}$ & $0.75$ & 0.89 \\
			$42447.26$ & $2.3 \times 10^{-10}$ & $3.57 \times 10^{-8}$ & $0.64$ & 0.79 \\
			$46663.69$ & $7 \times 10^{-11}$ & $6 \times 10^{-9}$ & $1$ & 1 \\
			$47767.504$ & $4.9 \times 10^{-10}$ & $8.10 \times 10^{-8}$ & $0.60$ & 0.67 \\
			$48945.6$ & $4.6 \times 10^{-11}$ & $4.2 \times 10^{-9}$ & $1.1$ & 1.3 \\
			$50020.04$ & $3 \times 10^{-11}$ & $2.1 \times 10^{-9}$ & $1$ & 2 \\
			$50260.031$ & $2.5 \times 10^{-10}$ & $3.19 \times 10^{-8}$ & $0.77$ & 1.1 \\
			$50458.94$ & $1.6 \times 10^{-10}$ & $6.1 \times 10^{-9}$ & $2.6$ & 3.0 \\
			$50489.7$ & $\dots$ & $8 \times 10^{-10}$ & $\dots$ & $\dots$  \\
			$50812.59$ & $8.8 \times 10^{-11}$ & $6.2 \times 10^{-9}$ & $1.4$ & 1.6 \\
			$51452.02$ & $1 \times 10^{-10}$ & $6.8 \times 10^{-9}$ & $1$ & 2 \\
			$51740.656$ & $4.1 \times 10^{-10}$ & $2.51 \times 10^{-8}$ & $1.7$ & 2.1 \\
			$51804.75$ & $7.6 \times 10^{-11}$ & $3.5 \times 10^{-9}$ & $2.2$ & 2.6 \\
			$52084.072$ & $3.0 \times 10^{-10}$ & $2.26 \times 10^{-8}$ & $1.3$ & 1.6 \\
			$52146.758$ & $8.1 \times 10^{-11}$ & $8.9 \times 10^{-9}$ & $0.91$ & 1.1 \\
			$52498.257$ & $1.0 \times 10^{-10}$ & $3.4 \times 10^{-9}$ & $2.9$ & 3.5 \\
			$52587.2$ & $7 \times 10^{-11}$ & $1.7 \times 10^{-9}$ & $4$ & 5 \\
			$53067.078$ & $8.8 \times 10^{-10}$ & $2.14 \times 10^{-7}$ & $0.41$ & 0.49 \\
			$53254.109$ & $3 \times 10^{-11}$ & $4.9 \times 10^{-9}$ & $0.6$ & 0.7 \\
			$53331.17$  & $1 \times 10^{-10}$ & $2.8 \times 10^{-9}$ & $4$ & 4  \\
			$53970.19$ & $4.4 \times 10^{-10}$ & $2.18 \times 10^{-8}$ & $2.0$ & 2.4 \\
			$54580.38$ & $3 \times 10^{-11}$ & $4.7 \times 10^{-9}$ & $1$ & 1 \\
			$55875.5$ & $\dots$ & $4.92 \times 10^{-8}$ & $\dots$ & $\dots$ \\
			$57839.92^\text{O2}$ & $3.9 \times 10^{-11}$ & $2.14 \times 10^{-9}$ & $1.8$ & 2.6 \\
			$58064.555$ & $9.9 \times 10^{-10}$ & $5.164 \times 10^{-7}$ & $0.19$ & 0.23 \\
			$58237.357$ & $6.6 \times 10^{-11}$ & $4.08 \times 10^{-9}$ & $1.6$ & 1.9 \\
			$58470.939$ & $5.1 \times 10^{-11}$ & $2.36 \times 10^{-9}$ &$2.2$ & 2.6 \\
			$58687.59^\text{O3}$ & $\dots$ & $3.60 \times 10^{-8}$ & $\dots$ & $\dots$ \\ [2pt]
			
			\hline\hline
		\end{tabular}
		
		&
		\hspace{-6pt}
		\begin{tabular}[t]{>{\centering}p{1.3cm} >{\centering}p{1.25cm} >{\centering}p{1.38cm} >{\centering}p{1.1cm} c} 
			\hline\hline \\[-8pt]
			\multicolumn{5}{c}{\small\textbf{Vela}} \\
			\hline \\[-6pt]
			\footnotesize MJD & \footnotesize $\varepsilon_{\text{ob}}(t_\text{g})$ & \footnotesize $\frac{\Delta\nu(t_\text{g})}{\nu_0}$ & \footnotesize $\frac{\varepsilon_{\text{ob}}(t_\text{g})\nu_0}{\Delta\nu(t_\text{g})}$[\%] & \footnotesize $\frac{\varepsilon_{\text{ob}}(t_\text{g})\nu_0}{Q\Delta\nu(t_\text{g})}$[\%] \\[4pt]
			
			$40280$ & $8.1 \times 10^{-9}$ & $2.34 \times 10^{-6}$ & $0.35$ & 10 \\
			$41192$ & $1.2 \times 10^{-8}$ & $2.05 \times 10^{-6}$ & $0.59$ & 17 \\
			$41312$ & $2 \times 10^{-9}$ & $1.2 \times 10^{-8}$ & $20$ & 40 \\
			$42683$ & $8.9 \times 10^{-9}$ & $1.99 \times 10^{-6}$ & $0.45$ & 2.1 \\
			$43693$ & $1.5 \times 10^{-8}$ & $3.06 \times 10^{-6}$ & $0.48$ & 4.0 \\
			$44888.4$ & $4.0 \times 10^{-8}$ & $1.145 \times 10^{-6}$ & $3.5$ & 20 \\
			$45192$ & $1.9 \times 10^{-8}$ & $2.05 \times 10^{-6}$ & $0.91$ & 21 \\
			$46257.228$ & $1.4 \times 10^{-8}$ & $1.601 \times 10^{-6}$ & $0.86$ & 5.5 \\
			$47519.8036$ & $6.2 \times 10^{-8}$ & $1.805 \times 10^{-6}$ & $3.5$ & 20 \\
			$48457.382$ & $4.9 \times 10^{-7}$ & $2.715 \times 10^{-6}$ & $18$ & 110 \\
			$49559$ & $0$ & $8.35 \times 10^{-7}$ & $0$ & 0 \\
			$49591.2$ & $9.7 \times 10^{-8}$ & $1.99 \times 10^{-7}$ & $49$  & 290 \\
			$50369.345$ & $4.8 \times 10^{-9}$ & $2.11 \times 10^{-6}$ & $0.23$ & 0.60\\
			$51559.319$ & $5.5 \times 10^{-9}$ & $3.086 \times 10^{-6}$ & $0.18$ & 1.0 \\
			$53193$ & $\dots$ & $2.1 \times 10^{-6}$ & $\dots$ & $\dots$ \\
			$53960$ & $1.9 \times 10^{-7}$ & $2.62 \times 10^{-6}$ & $7.1$ & 42 \\
			$55408.8$ & $6.1 \times 10^{-8}$ & $1.94 \times 10^{-6}$ & $3.1$ & 18 \\
			$56556$ & $1.2 \times 10^{-7}$ & $3.1 \times 10^{-6}$ & $3.9$ & 23 \\
			$56922$ & $8 \times 10^{-11}$ & $4 \times 10^{-10}$ & $20$ & 100 \\
			$57734.485^{\text{O2},1}$ & $5.9 \times 10^{-8}$ & $1.431 \times 10^{-6}$ & $4.2$ & 24 \\[2pt]
			
			\hline\hline
		\end{tabular}
	\end{tabular}
	
\end{table*}

We can see from Table \ref{tab:Percentage MoI contribution} that $\varepsilon_{\text{ob}}(t_\text{g})$ is typically $\sim$1\% of $\frac{\Delta\nu(t_\text{g})}{\nu_0}$ for the Crab, and typically 1 -- 4\% for Vela with the odd glitch hitting 20\% or even 50\%. Therefore, we find that for most glitches, the assumption of $\Delta I_{zz}$ being negligible holds at the moment of the glitch, as its contribution to the angular momentum has a relative size of $<5$\% when compared to the contribution due to $\Delta\nu$.  

We can extend this calculation to include the post-glitch recovery too. If we wanted to show that the change in the moment of inertia is negligible during the post-glitch recovery, we would need to show
\begin{align}
\frac{\Delta I_{zz}(t) }{I_{zz,0}} \ll \frac{\Delta \nu (t)}{\nu_0}~,
\end{align} 
for all $t>t_\text{g}$. However, $\Delta I_{zz}(t)$ at most has a value of $\Delta I_{zz}(t_\text{g})$ and the change in the spin frequency during the glitch recovery is typically $\Delta \nu (t) \sim \Delta \nu_\text{t}$. Therefore, we have 
\begin{align}
\label{MoI Condition 2}
\frac{\Delta I_{zz}(t_\text{g}) }{I_{zz,0}} \ll \frac{\Delta \nu_\text{t}}{\nu_0} = Q \left( \frac{\Delta \nu(t_\text{g})}{\nu_0} \right)~,
\end{align} 
where we used Equation~(\ref{unknownsToObservables}) in the equality. This condition is simply Equation~(\ref{MoI Condition}) but with an extra factor of $Q$. If $Q \sim 1$, then Equations~(\ref{MoI Condition 2}) and (\ref{MoI Condition}) are the same and since we have already shown Equation~(\ref{MoI Condition}) to be true, then Equation~(\ref{MoI Condition 2}) must also be true.

We do however, have values of $Q$ so we shall use them to calculate the ratio of the left hand side and the right hand side of Equation~(\ref{MoI Condition 2}). This ratio, expressed as a percentage, is presented in Table~\ref{tab:Percentage MoI contribution}. We see that the contribution from the moment of inertia is higher during the post-glitch recovery, though it is still small enough to allow us to justify neglecting it. There are a few Vela glitches which have a large contribution from the moment of inertia, but for the majority, it seems the inequality in Equation~(\ref{MoI Condition 2}) holds. Therefore, we can and we will ignore the change in the moment of inertia during the post-glitch recovery.

Our assumption of the change in the angular momentum being solely due to a change in spin frequency breaks down whenever the percentages in Table~\ref{tab:Percentage MoI contribution} get larger, like in a few of Vela's glitches. Nonetheless, we could still use the assumption but we would need to proceed with caution as a change in the moment of inertia about the rotation axis would contribute to a change in the NS's angular momentum. A quick way to parametrise how much the moment of inertia contributes to the change in angular momentum would be to look at the following parameters
\begin{align}
\eta_1 \equiv \frac{\varepsilon_{\text{ob}}(t_\text{g})\nu_0}{\Delta\nu(t_\text{g})} = - \frac{5}{32(2\pi)^4} \frac{c^5}{G} \frac{1}{I} \frac{\dot{\nu}_0}{\nu_0^5} \left(\frac{\Delta\dot{\nu}(t_\text{g})}{\dot{\nu}_0} \right) \left(\frac{\Delta\nu(t_\text{g})}{\nu_0} \right)^{-1},
\end{align}
\begin{align}
\eta_2 \equiv \frac{\varepsilon_{\text{ob}}(t_\text{g})\nu_0}{Q\Delta\nu(t_\text{g})} = - \frac{5}{32(2\pi)^4} \frac{c^5}{G} \frac{1}{I} \frac{\dot{\nu}_0}{\nu_0^5} \frac{1}{Q} \left(\frac{\Delta\dot{\nu}(t_\text{g})}{\dot{\nu}_0} \right) \left(\frac{\Delta\nu(t_\text{g})}{\nu_0} \right)^{-1},
\end{align} 
where $\eta_1$ represents the fractional contribution to the angular momentum due to a change in the moment of inertia at the glitch and $\eta_2$ represents the same but for the post-glitch recovery. If both these parameter are much less than 1, then we can ignore the change in the moment of inertia at the moment of the glitch and during the post-glitch recovery. If $\eta_1$ is considerably large, then it would mean there was a sudden increase in $\Delta I_{zz}$ at the glitch which would slow down the NS. This would mean whatever mechanism was producing the spin-up would have to be correspondingly larger than would otherwise be inferred, e.g. a larger starquake, a larger unpinning event. Similarly, a large $\eta_2$ would mean $\Delta I_{zz}$ is considerable during the recovery, i.e. the decaying mountain would spin-up the NS. To counteract the effect of including $\Delta I_{zz}$, the resultant transient mountain would need to be larger in size during the recovery.

The ratios $\eta_1$ and $\eta_2$ can be easily evaluated since the JBCA Glitch Catalogue and the ATNF Pulsar Catalogue provides all the information required. All in all, these two parameters are useful tools to assess whether the change in the moment of inertia about the rotation axis is an important factor when calculating a NS's dynamics at the moment of a glitch and during its subsequent recovery.


\bsp	
\label{lastpage}
\end{document}